\documentclass{mn2e}
\usepackage{psfig}
\def\simgt{\hbox{\rlap{\raise 0.425ex\hbox{$>$}}\lower 0.65ex\hbox{$\sim$}}}
\def\simlt{\hbox{\rlap{\raise 0.425ex\hbox{$<$}}\lower 0.65ex\hbox{$\sim$}}}

\def\bj {b_{\rm J}}

\def\kms {km$\,$s$^{-1}$}

\def\kmsmpc {km$\,$s$^{-1}$Mpc$^{-1}$}
\def\om {\Omega_{\rm m}}
\def\ol {\Omega_{\Lambda}}
\def \ho {H_0}

\def \mgii {Mg{\small~II}}

\def \civ {C{\small~IV}}
\def \ciii {C{\small~III}]}

\def \hb {H~$\beta$}
\def \fs {f_{\rm s}}
\def \fc {f_{\rm c}}
\def \nq {N_{\rm Q}}
\def \nobs {N_{\rm obs}}

\def \sqdeg {deg$^2$}
\def \isqdeg {deg$^{-2}$}
\def \ab {_{\rm AB}}
\def \sdss {_{\rm SDSS}}
\def \qso {_{\rm QSO}}
\def \gal {_{\rm gal}}
\def \qg {_{\rm qg}}
\def \zf {z_{\rm form}}
\def \mg {M_{\rm g}}

\def \all {_{\rm all}}
\def \bad {_{\rm bad}}

\def \mg {M_{\rm g}}

\def \aj {AJ}
\def \mnras {MNRAS}
\def \apj {ApJ}
\def \apjs {ApJS}

\def \aap {A\&A}

\title[The 2SLAQ QSO catalogue]{The 2dF-SDSS LRG and QSO Survey: The
  spectroscopic QSO catalogue.}

\author[S.~M. Croom et al.]
{Scott M. Croom$^{1,2}$\thanks{scroom@physics.usyd.edu.au},
Gordon T. Richards${^3}$,
Tom Shanks$^{4}$,
Brian J. Boyle$^{5}$,
\newauthor
Robert G. Sharp${^2}$,
Joss Bland-Hawthorn$^{1,2}$,
Terry Bridges${^2}$,
Robert J. Brunner$^{6}$,
\newauthor
Russell Cannon${^2}$,
Daniel Carson${^7}$,
Kuenley Chiu$^{8}$,
Matthew Colless$^2$,
\newauthor
Warrick Couch$^{9}$, 
Roberto De Propris$^{10}$,
Michael J. Drinkwater$^{11}$, 
Alastair Edge$^{4}$,
\newauthor
Stephen Fine${^1}$,
Jon Loveday$^{12}$,
Lance Miller$^{13}$,
Adam D. Myers$^{6}$,
Robert C. Nichol${^7}$,
\newauthor
Phil Outram$^{4}$, 
Kevin Pimbblet$^{11}$,
Isaac Roseboom$^{11,12}$,
Nicholas Ross$^{4,14}$,
\newauthor
Donald P. Schneider$^{14}$,
Allyn Smith$^{15}$,
Chris Stoughton$^{16}$,
Michael A. Strauss$^{17}$,
\newauthor
David Wake$^{4}$ 
\\
${^1}$ Institute of Astronomy, School of Physics, University of Sydney, NSW 2006, Australia\\
${^2}$ Anglo-Australian Observatory, PO Box 296, Epping, NSW 1710, 
Australia \\ 
${^3}$ Drexel University, Department of Physics, Philadelphia, PA
19104, USA\\
$^{4}$Department of Physics, University of Durham, South Road, Durham DH1 3LE\\
$^{5}$Australia Telescope National Facility, PO Box 76, Epping NSW 1710, Australia\\
$^{6}$Department of Astronomy, University of Illinois at Urbana-Champaign, Urbana, IL 61801\\
$^{7}$Institute of Cosmology and Gravitation, Mercantile House, Hampshire Terrace, University of Portsmouth, Portsmouth, PO1 2EG\\
$^{8}$School of Physics, University of Exeter, Stocker Road, Exeter EX4 4QL\\
$^{9}$Centre for Astrophysics \& Supercomputing, Swinburne University of Technology, P.O. Box 218, Hawthorn, VIC 3122, Australia\\
$^{10}$Cerro Tololo Inter-American Observatory, Casilla 603, La Serena, Chile\\
$^{11}$Department of Physics, University of Queensland, Brisbane, QLD 4072, Australia\\
$^{12}$Astronomy Centre, University of Sussex, Falmer, Brighton BN1 9QJ\\
$^{13}$Department of Physics, Oxford University, 1 Keble Road, Oxford,
OX1 3RH, UK\\
$^{14}$Department of Astronomy and Astrophysics, 525 Davey Laboratory,
	Pennsylvania State University, University Park, PA 16802.\\
$^{15}$Department of Physics and Astronomy, University of Wyoming,
	P.O. Box 3905, Laramie, WY 82071\\
$^{16}$Fermi National Accelerator Laboratory, P.O. Box 500, Batavia, IL 60510\\
$^{17}$Princeton University Observatory, Peyton Hall, Princeton, NJ
	08544, USA
}

\begin{document}

\maketitle

\newcommand{\fmmm}[1]{\mbox{$#1$}}
\newcommand{\scnd}{\mbox{\fmmm{''}\hskip-0.3em .}}
\newcommand{\scnp}{\mbox{\fmmm{''}}}

\begin{abstract}

We present the final spectroscopic QSO catalogue from the 2dF-SDSS LRG
and QSO (2SLAQ) Survey.  This is a deep, $18<g<21.85$ (extinction
corrected), sample aimed at probing in detail the faint end of the
broad line AGN luminosity distribution at $z\simlt2.6$. The candidate QSOs
were selected from SDSS photometry and observed spectroscopically with
the 2dF spectrograph on the Anglo-Australian Telescope.  This sample
covers an area of 191.9 deg$^2$ and contains new spectra of 16326
objects, of which 8764 are QSOs, and 7623 are newly discovered (the
remainder were previously identified by the 2QZ and SDSS surveys).  The full QSO
sample (including objects previously observed in the SDSS and 2QZ
surveys) contains 12702 QSOs.  The new 2SLAQ spectroscopic data set
also contains 2343 Galactic stars, including 362 white dwarfs, and 2924
narrow emission line galaxies with a median redshift of $z=0.22$.

We present detailed completeness estimates for the survey, based on
modelling of QSO colours, including host galaxy contributions.  This
calculation shows that at $g\simeq21.85$ QSO colours are significantly
affected by the presence of a host galaxy up to redshift $z\sim1$ in the SDSS
$ugriz$ bands.  In particular we see a significant reddening of the
objects in $g-i$ towards fainter $g$-band magnitudes.  This reddening
is consistent with the QSO host galaxies being dominated by a stellar
population of age at least $2-3$~Gyr.

The full catalogue, including completeness estimates, is available
on-line at {\tt http://www.2slaq.info/}.

\end{abstract}


\section{Introduction}

The last decade has seen the coming of age of extremely high multiplex
fibre spectroscopy, as implemented by the 2-degree Field (2dF)
instrument \cite{2dfpaper} and the Sloan Digital Sky Survey (SDSS; York
et al. 2000).  These new facilities have
allowed order of magnitude increases in sample sizes over the previous
generation of surveys.  The 2dF QSO Redshift Survey (2QZ;
Croom et al. 2001a; 2004) and the SDSS QSO survey (Schneider et al. 2007)
have allowed precise measurement of the evolution of QSOs (e.g. Boyle
et al. 2000; Croom et al. 2004; Richards et al. 2006), QSO
clustering (e.g. Croom et al. 2001b; Croom et al. 2005; Shen et
al. 2007), spectral properties (e.g. Croom et al. 2002; Corbett et
al. 2003; Vanden Berk et al. 2001; Richards et al. 2002a) and a range
of other significant results.  The published sample sizes
($\sim25000$ QSOs in 2QZ; $\sim80000$ QSOs in SDSS) are large enough
that in many cases measurements are now limited by systematic
uncertainties rather than random errors. 

However, one of the important limitations of the 2QZ and SDSS surveys
are their relative depths.  The SDSS QSO survey is limited to $i=19.1$,
or $i=20.2$ for the high redshift sample \cite{r02b}, which does not
reach the break in the QSO luminosity function.  2QZ is somewhat deeper, limited
in the bluer $\bj$-band to $\bj<20.85$.  The 2QZ clearly shows the
break in the QSO luminosity function (LF), typically reaching $\sim1$
mag fainter than the break at $z<2$.  The observed break in the LF is
a gradual flattening towards faint magnitudes; as a result
the constraints from the 2QZ on the actual slope of the faint end are
fairly uncertain, as evidenced by the difference between the results
from the first release (Boyle et al. 2000) and the final release (Croom
et al. 2004; C04).  In
comparison, X-ray surveys, in particular using {\it Chandra}
(e.g. Giacconi et al. 2002; Alexander et al. 2003) and {\it XMM-Newton}
(e.g. Hasinger et al. 2001; Worsley et al. 2004; Barcons et al. 2007), reach to fainter
depths, but over a much smaller area.  The largest samples contain
$\sim1000$ objects over a few square degrees.  These surveys have
demonstrated that the pure luminosity evolution that appears to model
the evolution of the most extensive optical samples (e.g. 2QZ, SDSS) fails to trace
the evolution of the faint AGN populations at $L<L^*$.  It now
appears that the activity in faint AGN peaks at a lower redshift than
that of more luminous AGN (e.g. Hasinger et al 2005);  this process
has been described as AGN downsizing (e.g. Barger et al. 2005).
Whether the downsizing is due to lower mass black holes being more
active at low redshift (e.g. Heckman et al. 2004) or massive black
holes at lower rates of accretion (e.g. Babic et al., 2007) remains
unclear.  Both effects are likely to play a role.

Substantial advances have been made in the theoretical understanding
of AGN formation and the connection to galaxy formation (e.g. Hopkins
et al. 2005a).  This work has been largely driven by the observational
evidence that most massive galaxies with bulges contain super-massive BHs (SMBHs)
(e.g. Tremaine et al. 2002).  SMBH accretion is thought to be
triggered (at least for moderate to high luminosity AGN) by the merger
of gas rich galaxies; while the timescale for the merger may be as
long as $\sim1$Gyr, during the majority of this time the accretion is obscured from
view.  It is only when the AGN finally expels the surrounding gas and
dust that it shines like a quasar for a brief period ($\sim100$Myr),
before exhausting its fuel supply (e.g. Di Matteo et al. 2005).  This
feedback of the AGN into the host also heats (and possibly expels) the
gas in the galaxy, which suppresses star formation 
leading to ``red and dead'' ellipticals or bulges.  These models match
a number of previous observations and predict that the faint end of
the QSO luminosity function is largely  comprised of higher mass BHs
at lower accretion rates (i.e. below their peak luminosity) \cite{hop05b}.

The 2dF SDSS LRG and QSO (2SLAQ) Survey was designed to survey
optically faint AGN/QSOs within a sufficiently large volume to obtain
robust measurements of both the luminosity function and QSO
clustering.  Throughout this paper we will use the term QSO to refer
to any broad line (type 1) AGN, irrespective of luminosity.  The QSO
portion of the survey shared 
fibres with a related program to target luminous red galaxies (LRGs)
at $z\simeq0.4-0.7$ (Cannon et al. 2006).  Both the LRGs and QSOs were
selected from single epoch SDSS imaging data, and then observed
spectroscopically with the 2-degree Field (2dF) instrument at the
Anglo-Australian Telescope (AAT).  The 2SLAQ QSO sample has already
produced a preliminary QSO luminosity function (Richards
et al. 2005; R05), measured the clustering of QSOs as a function of
luminosity (da Angela et al. 2008) and studied the distribution of QSO
broad line widths (Fine et al. 2008).  In this paper we present the
final spectroscopic QSO catalogue of the 2SLAQ sample.  We then carry
out a detailed analysis of the survey completeness.  The analysis of
the QSO luminosity function from the final 2SLAQ sample is presented in a
companion paper (Croom et al. in prep).

In Section \ref{sec:imaging} we discuss the selection of QSO
candidates from the SDSS imaging data.  This has largely been
described by R05, but is summarized here for
completeness.  In Section \ref{sec:spec} we present the spectroscopic
observations, and in Section \ref{sec:cat} we describe the composition
and quality of the resulting catalogue.  Section \ref{sec:comp}
contains our detailed completeness analysis.  We summarize our
results in Section \ref{sec:sum}.  Throughout this paper we will
assume a cosmology with $\ho=70$\kmsmpc, $\om=0.3$ and $\ol=0.7$.

\section{Imaging data and QSO selection}\label{sec:imaging}

\subsection{The SDSS imaging data}

The photometric measurements used as the basis for our catalogue are
drawn from the Data Release 1 (DR1) processing (Stoughton et~al. 2002;
Abazajian et~al. 2003) of the SDSS imaging data.  The astrometric
precision at the faint limit of the survey is $\sim0.1$ arcsec
\cite{pier03}.  The SDSS data are
taken in five photometric pass-bands ($ugriz$; Fukugita et al. 1996)
using a large format CCD camera \cite{gunn98} on a special-purpose
2.5-m telescope (Gunn et al. 2006).  The regions covered by
the 2SLAQ survey were complete in DR1, so no further updates to more
recent data releases are required.  Except where otherwise
stated, all SDSS magnitudes discussed herein are ``asinh''
point-spread-function (PSF) magnitudes \cite{lupton99} on the SDSS
pseudo--AB magnitude system \cite{og83} that have been dereddened
for Galactic extinction according to the model of Schlegel, Finkbeiner
\& Davis (1998).  The SDSS Quasar Survey \cite{sdssqso4} extends to
$i=19.1$ for $z<3$ and $i=20.2$ for $z>3$; our work herein explores
the $z<3$ regime to $g=21.85$ ($i\sim21.63$, based on the median
colours of SDSS QSOs, e.g. Richards et al. 2001).  At the faint limit of
the 2SLAQ sample ($21.75<g<21.85$), the photometric errors are
typically $\Delta u=0.20$, $\Delta g=0.07$, $\Delta r=0.08$, $\Delta
i=0.11$, $\Delta z=0.33$.

\subsection{Sample selection}\label{sec:sel}

\subsubsection{Preliminary sample restrictions}

Our quasar candidate sample was drawn from 10 SDSS imaging runs.  Due
to the slightly poorer image quality in the 6th row 
of CCDs in the SDSS camera we did not include these data.  Thus the
2SLAQ survey regions are $2^\circ$ wide rather
  than the usual $2^\circ.5$ for an SDSS imaging run.  We rejected
    any objects that met the 
``fatal'' or ``non-fatal'' error definitions of the SDSS quasar target
selection \cite{r02a}.  Although
our survey covers the southern equatorial Stripe 82 region which has
been scanned multiple times \cite{am08}, the co-added data (Annis et al. 2006)
were not available at the time of our spectroscopic observations and
so single scan data were used.

We apply a limit to the (extinction corrected) $i$-band PSF magnitude
of $i<22.0$ and $\sigma_i<0.2$.  We also placed restrictions on the
errors in each of the other four bands: 
$\sigma_u<0.4$, $\sigma_g<0.13$, $\sigma_r<0.13$ and
$\sigma_z<0.6$.  Note that this selection of error
constraints effectively limits the redshift to less than 3, as the
Ly$\alpha$ forest suppresses the $u$ flux at higher redshifts.

\subsubsection{Low redshift colour cuts}

Based on spectroscopic identifications from SDSS and 2QZ of this
initial set of objects, we implement additional colour cuts that are
designed to select faint UV-excess QSOs with high efficiency and
completeness at redshifts $z\simlt2.6$.  An
analysis of the completeness of the selection algorithm is given as a
function of redshift and magnitude in Section \ref{sec:photcomp}.

We reject hot white dwarfs using the following cuts, independent of
magnitude.  Specifically, we rejected objects that satisfy the condition:
A AND ((B AND C AND D) OR E), where the letters refer to the cuts:
\begin{eqnarray}
\begin{array}{rrcccl}
{\rm A}) \;\, & -1.0 & < & u-g & < & 0.8 \\
{\rm B}) \;\, & -0.8 & < & g-r & < & 0.0 \\
{\rm C}) \;\, & -0.6 & < & r-i & < & -0.1 \\
{\rm D}) \;\, & -1.0 & < & i-z & < & -0.1 \\
{\rm E}) \;\, & -1.5 & < & g-i & < & -0.3. 
\end{array}
\end{eqnarray}
These constraints are similar to the white dwarf cut applied by Richards
et~al. (2002a; their Eq. 2) except for the added cut with respect to 
the $g-i$ colour.

\begin{figure}
\centering
\centerline{\psfig{file=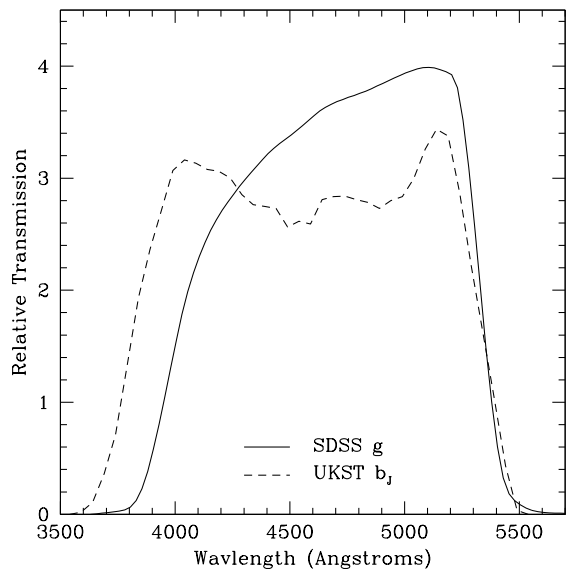,width=8cm}}
\caption{The UKST $\bj$ (Maddox \& Hewett 2006) and SDSS $g$
  transmission curves (detector quantum efficiency) for airmass of 1.
  Normalization is such that $\int S_\lambda {\rm
  d}\lambda/\lambda=1$.  This is a corrected version of the same plot
  shown by R05, in which the conversion from energy to photon
  efficiency for the $\bj$ response was incorrect.} 
\label{fig:bjg}
\end{figure}

To efficiently target both bright and faint targets we use different
colour cuts as a function of $g$-band magnitude.  The bright sample is
restricted to $18.0<g<21.15$ and is designed to allow for overlap with
previous SDSS and 2dF spectroscopic observations.  The faint sample
has $21.15\le g<21.85$ and probes roughly one magnitude deeper than
2QZ.  These cuts are made in $g$, rather than the $i$-band that the
SDSS quasar survey uses, since we are concentrating on UV-excess
quasars and would like to facilitate comparison with the results from
the $\bj$-based 2QZ.  At this depth an $i$-band limited sample
selected from single epoch SDSS data would also contain substantial
stellar contamination.  The combination of the $g<21.85$ and $i<22.0$
cuts will exclude objects bluer than $\alpha_{\nu}=+0.3\;
(f_{\nu}\propto\nu^{\alpha})$; however, objects this blue are
exceedingly rare ($>3\sigma$ deviations from the observed QSO spectral
slope distribution; Richards et al 2006).  As
pointed out by R05, the $\bj$ and $g$ bands are almost equivalent,
with a mean $(g-\bj)=-0.045$ found for QSOs in common between the SDSS
and 2QZ.  We note that the relative
transmission curves plotted in R05 (their Fig. 6) were in error (the
energy to photon conversion was reversed).  We plot the correct
comparison in Fig. \ref{fig:bjg}.

In general, we would prefer to avoid a
morphology-based cut since we do not want to exclude low-$z$ quasars
and because our selection extends beyond the magnitude limits at which
the SDSS's star/galaxy separation breaks down.  However,
Scranton et~al. (2002) have developed a Bayesian star-galaxy classifier that
is robust to $r\sim22$.  As a result, in addition to straight
colour-cuts, we also apply some colour restrictions on objects with
high $r$-band galaxy probability (referred to below as ``galprob'')
according to Scranton et~al. (2002) in an attempt to remove contamination from
narrow emission line galaxies (NELGs; i.e. blue star-forming
galaxies) from our target list.

Bright sample objects are those with $18.0<g<21.15$ and that meet the
following conditions
\begin{eqnarray}
\begin{array}{rcclcccll}
{\rm A}) \;\, & u-g & < & 0.8 & {\rm AND} & g-r & < & 0.6 & {\rm AND} \\
 & r-i & < & 0.6 & & & & & \\
{\rm B}) \;\, & u-g & > & 0.6 & {\rm AND} & g-i & > & 0.2 & \\
{\rm C}) \;\, & u-g & > & 0.45 & {\rm AND} & g-i & > & 0.35 & \\
{\rm D}) \;\, & {\rm galprob} & > & 0.99 & {\rm AND} & u-g & > & 0.2 &
 {\rm AND} \\
 & g-r & > & 0.25 & {\rm AND} & r-i & < & 0.3 & \\
{\rm E}) \;\, & {\rm galprob} & > & 0.99 & {\rm AND} & u-g & > & 0.45. &
\end{array}
\end{eqnarray}
in the combination A AND (NOT B) AND (NOT C) AND (NOT D) AND
(NOT E), where cut A selects
UVX objects, cuts B and C eliminate faint F-stars whose metallicity
and errors push them blueward into the quasar regime, and cuts D and E
remove NELGs that appear extended in the $r$ band.  Among the bright
sample objects, those with $g>20.5$ were given priority in terms of
fibre assignment (see Section \ref{sec:targpri}).

Faint sample objects are those with $21.15\le g<21.85$ and that meet the
following conditions
\begin{eqnarray}
\begin{array}{rcclcccll}
{\rm A}) \;\, & u-g & < & 0.8 & {\rm AND} & g-r & < & 0.5 & {\rm AND} \\
 & r-i & < & 0.6 & & & & & \\
{\rm B}) \;\, & u-g & > & 0.5 & {\rm AND} & g-i & > & 0.15 \\
{\rm C}) \;\, & u-g & > & 0.4 & {\rm AND} & g-i & > & 0.3 \\
{\rm D}) \;\, & u-g & > & 0.2 & {\rm AND} & g-i & > & 0.45 \\
{\rm E}) \;\, & {\rm galprob} & > & 0.99 & {\rm AND} & g-r & > & 0.3.
\end{array}
\end{eqnarray}
in the combination A AND (NOT B) AND (NOT C) AND (NOT D) AND (NOT E),
where cut A selects UVX objects, cuts B, C and D eliminate faint F-stars whose
metallicity and errors push them blueward into the quasar regime, and
cut E removes NELGs.  These faint cuts are more restrictive than the
bright cuts to avoid significant contamination from main sequence
stars that will enter the sample as a result of larger errors at
fainter magnitudes.  The low redshift colour cuts ($u-g$ and
$g-i$) are shown in Fig. \ref{fig:ugi_gz} (see also Fig. 1 of R05).

\subsubsection{High redshift colour cuts}\label{sec:highzsel}

In addition to the main low redshift ($z\simlt2.6$) sample described
above, we also target a sample of higher redshift QSO candidates,
analogous to the high redshift sample selected in the main SDSS QSO
survey which selected QSOs up to $z\simeq5.4$ at $i<20.2$ (Richards et
al. 2002a).  The 2SLAQ high redshift sample was limited to $i<21.0$
and an additional constraint that $\sigma_z<0.4$ was applied to the
$z$-band photometry.  We then selected candidates in three redshift
intervals.  QSO candidates at redshift $\simeq3.0-3.5$ satisfied the
following cuts:
\begin{eqnarray}
\begin{array}{cclc}
\sigma_r & < & 0.13 & {\rm AND} \\
u        & > & 20.6 & {\rm AND} \\
u-g      & > & 1.5  & {\rm AND} \\
g-r      & < & 1.2  & {\rm AND} \\
r-i      & < & 0.3  & {\rm AND} \\
i-z      & > & -1.0 & {\rm AND} \\
g-r      & < & 0.44(u-g)-0.76.\\
\end{array}
\end{eqnarray}
For the redshift range $\simeq3.5-4.5$ this selection becomes
\begin{eqnarray}
\begin{array}{ccclccclc}
{\rm A}) \;\, &\sigma_r & < & 0.2 & \\
{\rm B}) \;\, &u-g      & > & 1.5  & OR & u & > & 20.6 \\
{\rm C}) \;\, &g-r      & > & 0.7  & \\
{\rm D}) \;\, &g-r      & > & 2.8  & OR & r-i & < & 0.44(g-r)-0.558\\
{\rm E}) \;\, &i-z      & < & 0.25 & {\rm AND} & i-z & > & -1.0,\\
\end{array}
\end{eqnarray}
%
in the combination A AND B AND C AND D AND E.  For the redshifts above
$\simeq4.5$ we use 
\begin{eqnarray}
\begin{array}{cclc}
u        & > & 21.5 & {\rm AND} \\
g        & > & 21.0  & {\rm AND} \\
r-i      & > & 0.6  & {\rm AND} \\
i-z      & > & -1.0 & {\rm AND} \\
i-z      & < & 0.52(r-i)-0.762.\\
\end{array}
\end{eqnarray}
%
These samples have a high degree of contamination from the stellar
locus due to photometric errors.  These candidates were therefore
targeted at a lower priority than the main low redshift sample, and we
do not present a detailed analysis of completeness for the high
redshift sample.

\subsection{Survey area}

\begin{table}
\begin{center}
\caption{Coordinates of the 2SLAQ survey regions.}
\setlength{\tabcolsep}{4pt}
\begin{tabular}{cccccccccc}
\hline
2SLAQ  & \multicolumn{2}{c}{RA (J2000)} & \multicolumn{2}{c}{Dec (J2000)} \\
Region & min & max & min & max\\
\hline
a & 123.0 & 144.0 & -1.259 & 0.840\\
b & 150.0 & 168.0 & -1.259 & 0.840\\
c & 185.0 & 193.0 & -1.259 & 0.840\\
d & 197.0 & 214.0 & -1.259 & 0.840\\
e & 218.0 & 230.0 & -1.259 & 0.840\\
s & 309.0 & 59.70 & -1.259 & 0.840\\
\hline
\label{tab:surv_regions}
\end{tabular}
\end{center}
\end{table}

\begin{table}
\begin{center}
\caption{SDSS imaging runs used for 2SLAQ target selection.  We list
  the run number, modified Julian date (MJD) of observation and the 2SLAQ
  regions that each run contributes to.  Note that runs can contribute
  to more than one 2SLAQ region.} 
\setlength{\tabcolsep}{4pt}
\begin{tabular}{ccc}
\hline
SDSS run & MJD & 2SLAQ regions\\
\hline
752  & 51258 & c, d, e\\
756  & 51259 & a, b, c, d, e\\
1239 & 51607 & a\\
2141 & 51962 & b\\
2583 & 52172 & s\\
2659 & 52197 & s\\
2662 & 52197 & s\\
2738 & 52234 & s\\
3325 & 52522 & s\\
3388 & 52558 & s\\
\hline
\label{tab:runs}
\end{tabular}
\end{center}
\end{table}

The survey was targeted along the two equatorial regions from the SDSS
imaging data.  In the North Galactic Cap, we selected five disjoint
regions along $\delta\simeq0^\circ$ which contain the best quality
imaging data.  These are denoted as regions a, b, c, d, e, as listed
in Table \ref{tab:surv_regions}.  In the South Galactic Cap we
targeted a single contiguous region, denoted as s.  The 10 SDSS
imaging runs used are listed in Table \ref{tab:runs}, along with the
2SLAQ regions to which they contribute.  The 2SLAQ area completely
overlaps with the brighter SDSS QSO survey (e.g. Schneider et
al. 2007).  There is partial overlap with the 2QZ (C04)
in the North Galactic Cap, with the 2QZ covering the RA range
$148^\circ<\alpha_{\rm J2000}<223^\circ$.

\section{Spectroscopic observations}\label{sec:spec}

\subsection{Instrumental setup}

Spectroscopic observations of the input catalogue were made with the
2-degree Field (2dF) instrument at the Anglo-Australian Telescope
\cite{2dfpaper}.  The 2dF instrument is a multi-fibre spectrograph which
can obtain simultaneous spectra of up to 400 objects over a
$2^{\circ}$ diameter field of view, and is located at the prime
focus of the telescope.  Fibres are robotically
positioned within the field of view and are fed to two identical
spectrographs (200 fibres each).  Two field plates and a tumbling
system allow one field to be observed while a second is being
configured, reducing down-time between fields to a minimum.  The
spectrographs each contain a Tektronix $1024\times1024$ CCD with
24~$\mu$m pixels.

Observations of QSOs and LRGs were combined by using 200 fibres for
each sample and sending these to separate spectrographs.  QSO targets
were sent to spectrograph 1 which contained a low resolution 300B grating
with a central wavelength of 5800\AA.  LRG targets were directed to
spectrograph 2 with a higher resolution 600V grating centred at
6150\AA\ (see Cannon et al. 2006 for further details of the LRG
sample). The 300B grating produces a dispersion of 4.3\AA~pixel$^{-1}$,
giving an instrumental resolution of 9\AA.  The spectra covered the
wavelength range 3700--7900\AA.

\subsection{Target configuration and priority}\label{sec:targpri}

\begin{table}
\begin{center}
\caption{Configuration priorities for 2SLAQ targets.  9 is the highest
priority, while 1 is the lowest.}\label{tab:pri}
\begin{tabular}{lc}
\hline
Sample & Priority \\
\hline
Guide stars               & 9 \\
LRG (main) random         & 8 \\
LRG (main) remainder      & 7 \\
QSOs ($g>20.5$) random    & 6 \\
QSOs ($g>20.5$) remainder & 5 \\
LRG(extras)+hi-z QSOs     & 4 \\
QSOs ($g<20.5$)           & 3 \\
previously observed       & 1 \\
\hline
\end{tabular}
\end{center}
\end{table}

The 2dF {\small CONFIGURE} program \cite{configure} was used to
allocate specific fibres to objects.  This software takes an input
list  of prioritized positions (including guide fibres and target
positions) and through an iterative scheme allocates fibres, producing
a second file which is passed to the control software for the 2dF
robotic positioner.  For the 2SLAQ observing program, minor
modifications to the {\small CONFIGURE} software were made to allow i)
fibres from different spectrographs to be allocated to different
samples, ii) different central wavelengths for each spectrograph.  We
also carried out a detailed analysis of the spatial variation of
configured target density across the 2dF field.  This showed that the
algorithm could, in certain circumstances, impart considerable
structure on the distribution of targets.  The main effects seen were
a deficit of objects near the centre of the field ($<0.25^{\circ}$
radius) in high density fields (where the number of targets is greater
than the number of fibres) and systematics related to the ordering of
targets.  To address these issues, the targets were randomized and
randomly re-sampled so that so that the highest priority targets had a
surface density of 70 deg$^{-2}$.  We note that these issues have
since been fully addressed by Mizarksi et al. (2006) using a simulated
annealing algorithm; however, 2SLAQ observations were carried out
prior to this work.

Our most important targets were given higher priority in the fibre
configuration process.  These priorities are summarized in Table
\ref{tab:pri}.  LRGs were given highest priority because they have a
lower surface density than the QSO candidates.  Our highest priority
QSO targets had $g>20.5$ and were given a priority of  6.  The surface
density of these targets was significantly higher than the $\sim70$
deg$^{-2}$ that can be configured with the available fibres and so
were randomly re-sampled.  The remaining $g>20.5$ QSOs (not selected
in the random sampling) had their priority set to 5.  The high
redshift QSO candidates had their priorities set to 4, and the bright
QSO candidates ($g<20.5$) had priority of 3.  If a 2SLAQ selected
source already had a high quality spectroscopic observation from
either 2QZ or SDSS its priority was set to 1 (lowest on a scale of
$1-9$) in the 2dF configuration (i.e. it was observed only if no other
target was available).

\subsection{Tiling of 2dF fields}

\begin{figure*}
\centering
\centerline{\psfig{file=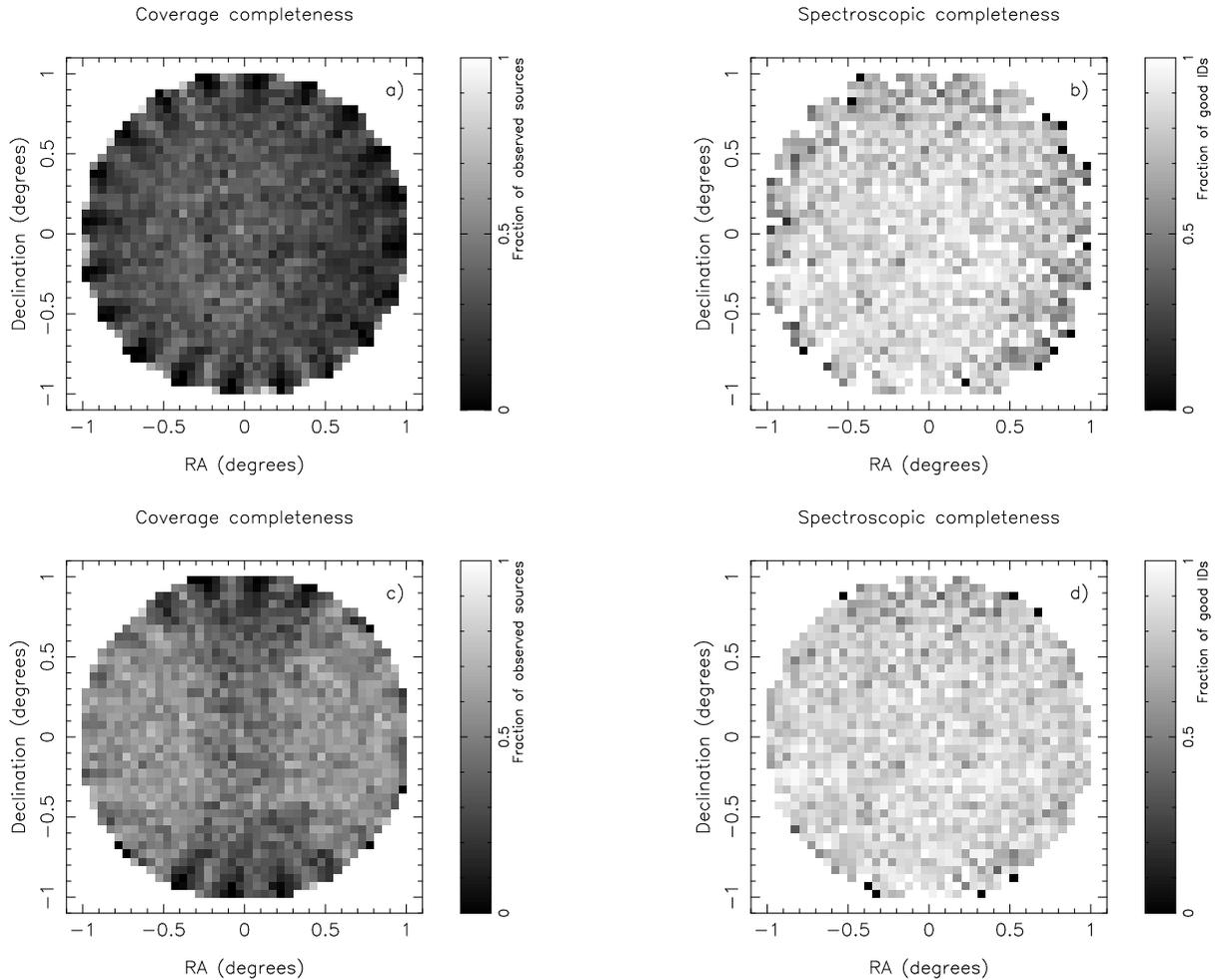,width=16cm,angle=270}}
\caption{2SLAQ survey completeness as a function of 2dF field
  position. a) Average coverage completeness within
  individual 2dF pointings.  Note the small triangular regions at the
  edge of the field which are inaccessible due to alternating
  blocks of 10 fibres going to the LRG sample.  There is also a slight
  deficit of observed targets on the right hand side of the field,
  due to the sky fibres preferentially coming from that side of the
  field.  This is compensated for by overlapping fields. b) Average
  spectroscopic completeness for individual 2dF pointings.  A small
  decline is visible towards the field edge (see
  Fig. \ref{fig:radcomp}). c) Coverage completeness for 
  all objects within a 2dF field radius (allowing for overlaps).
  Regions of increased completeness due to overlapping fields can be
  seen to the East and West edges of the field. d) Spectroscopic completeness  for
  all objects within a 2dF field radius (allowing for overlaps).  This
  distribution is quite uniform over the entire field.}
\label{fig:fldcomp}
\end{figure*}

\begin{table}
\begin{center}
\caption{2SLAQ observed objects which are outside of the nominal
  survey limits bounded by the intersection of $1.05^\circ$ radius 2dF
  fields of view.}\label{tab:outside}
\begin{tabular}{ccc}
\hline
Name & RA (J2000) & Dec. (J2000) \\
     & (deg)      & (deg)        \\
\hline
J005128.22$+$004447.8 &  12.867619 &   0.746637 \\
J010159.56$+$004820.2 &  15.498189 &   0.805634 \\
J010423.79$+$004029.8 &  16.099131 &   0.674965 \\
J022526.32$-$011434.3 &  36.359680 &  -1.242874 \\
J081233.10$+$004643.0 & 123.137909 &   0.778604 \\
J081238.23$+$004713.2 & 123.159286 &   0.786992 \\
J081938.22$-$011052.6 & 124.909256 &  -1.181273 \\
J100705.00$-$010904.9 & 151.770844 &  -1.151354 \\
J123158.37$+$004635.6 & 187.993195 &   0.776566 \\
J123448.89$+$004752.3 & 188.703705 &   0.797856 \\
J134054.18$+$004911.9 & 205.225769 &   0.819972 \\
J143313.87$-$011501.2 & 218.307800 &  -1.250338 \\
J143620.20$+$004529.6 & 219.084152 &   0.758220 \\
J143922.06$-$011215.0 & 219.841934 &  -1.204170 \\
J144007.84$+$004156.2 & 220.032654 &   0.698939 \\
J211844.76$+$003134.4 & 319.686523 &   0.526249 \\
J211955.31$+$004301.6 & 319.980469 &   0.717123 \\
J212129.68$+$004827.1 & 320.373688 &   0.807553 \\
J214859.57$+$004439.5 & 327.248230 &   0.744307 \\
\hline
\end{tabular}
\end{center}
\end{table}

Given the geometry of the imaging area (strips between 10$^\circ$ and
110$^\circ$ long, which are all $2^{\circ}$ wide), it was
sensible to employ a simple tiling pattern to cover
the 2SLAQ regions.  Each circular 2dF field was spaced along the strip
at intervals of $1.2^\circ$ in RA.  In some cases field centres were
shifted slightly to optimize their positions (e.g. at the end of
survey regions).  Some of the first fields observed (February - April 2003)
had a smaller spacing of $1^\circ$.  The 1.2$^\circ$ spacing produced a near
optimal balance between coverage and completeness for both the QSOs
and LRGs.  This approach also provides some overlap between adjacent fields.  The
2dF field of view has a radius of $1.05^\circ$.  For most observations,
configured objects were constrained to lie within a circle of exactly
$1.05^\circ$ radius; however, early observations did not apply this
constraint (February -- September 2003).  As a result, 19 2SLAQ
objects (including 10 good quality QSOs) were observed marginally
outside the nominal survey region bounded by the intersection of
all the observed fields each with radius $1.05^\circ$.  This can be
caused by effects such as atmospheric refraction which distorts the
field of view at high airmass.  These are
included in the catalogue, but excluded in several of the analyses
below; the names of these sources are given in Table
\ref{tab:outside}.

\begin{figure*}
\centering
\centerline{\psfig{file=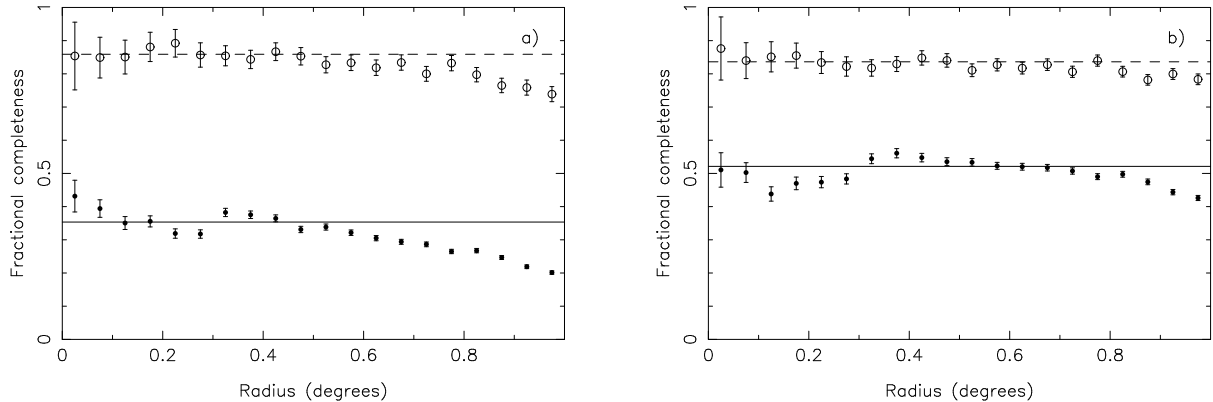,width=16cm,angle=270}}
\caption{2SLAQ survey completeness as a function of radius from 2dF
  field centres averaged over all 2SLAQ fields. a) Coverage (filled
  circles) and 
  spectroscopic (open circles) completeness within individual 2dF
  pointings. A clear decline towards the field edge is seen in both
  cases. The solid and dashed lines show the mean coverage and
  spectroscopic completeness respectively within a radius of
  $0.5^\circ$. b) As for a), but calculated for all objects within a
  2dF field radius (allowing for overlaps).  The depression at less
  than $\sim0.3^\circ$ in the coverage (filled points) is because the
  overlapping fields do not quite reach to the middle of the adjacent
  field (c.f. Fig. \ref{fig:fldcomp}).}
\label{fig:radcomp}
\end{figure*}

In order to maximize the yield from overlapping fields, sequences of
alternating fields were generally observed first, with the interleaved
overlapping fields observed in a second pass.  On this second pass,
all targets which obtained high quality identifications (quality 1;
see Section \ref{sec:specid} below) from previous observations
were given the lowest priority (priority 1), so that a minimal number
of objects with acceptable spectroscopic data were repeated.  Even
allowing for this, 3317 objects have repeat
observations.  Approximately half of these were because the original
spectrum was low quality.  The other half were repeated because there
was no higher priority target accessible.  These repeated spectra are
useful in making internal checks of completeness and consistency.

There are also a number of physical constraints on the configuration
of 2dF fields.  In the 2SLAQ survey the most apparent of these is that
fibres are arranged around the edge of the field plates in blocks of
10.  These blocks of 10 fibres go to alternating spectrographs, such
that there is a triangular region directly in front of each fibre
block going to spectrograph 2 that fibres from spectrograph 1 cannot
access.  This is because 2dF fibres are limited to a maximum
off-radial angle of $14^{\circ}$.  The 20 small inaccessible triangles
amount to a total area of 0.43 deg$^2$.  As a number of different
samples are configured together in each field, the distribution of
other targets also influences the angular selection function within
2dF fields.  The main QSO sample was given lower priority than the
main LRG sample, so great care needs to be taken determining
statistics (e.g. clustering) which depend on the angular distribution
of QSOs (see da Angela et al. 2008).  In addition, 2dF fibres cannot be
positioned closer than $\sim30$ arcsec, which can reduce the number of
close pairs on these angular scales.  Finally, 20 fibres were allocated to
sky positions.  Each 2dF spectrograph CCD takes data from 200 fibres.
The sky positions were allocated to fibres that lay in the central 100
fibres on the spectrograph CCD (which predominantly come from the western
side of the 2dF field).  This region on the CCD has the best spectral
and spatial PSF, allowing PSF mapping and convolution to improve sky
subtraction if required \cite{willis01}.

The influence of these varied effects is displayed in
Fig. \ref{fig:fldcomp}.  The unreachable triangular regions near the
edge of the field are clearly visible in Fig. \ref{fig:fldcomp}a,
while when overlaps are considered (Fig. \ref{fig:fldcomp}c) these
features are only seen at the very top and bottom of the field.
Radial variations in spectroscopic completeness are also visible
(Fig. \ref{fig:fldcomp}b; Fig. \ref{fig:radcomp}).  Both coverage and
spectroscopic completeness gradients are visible, but these are much
less pronounced when the overlaps between fields are taken into
account.

\subsection{2dF observations}

2SLAQ observations were carried out over a period February 2003 to
August 2005, using a total of 89 nights of AAT time.  The fiducial
exposure time for each field was 4 hours.  Due to the effects of
differential spatial atmospheric refraction across the 2 degree field
of view, a single field could not be observed for more than $\sim2$
hours at a time (and significantly less if observed at high airmass),
so a field would typically be observed over 2 nights, with 4x1800s
exposures being taken each night.

Data reduction and quality assessment at the telescope enabled
determination of whether the nominal spectroscopic completeness  
limit for QSO candidates had been obtained ($>80$ per cent quality 1
IDs; see Section \ref{sec:specid} below).
Further observations were taken if this limit was not obtained,
usually because of poor weather.  This
analysis allowed us to identify those objects which had
sufficient $S/N$ for a good identification in only the first 2 hours
of observing a field.  Any fibres on such objects were
re-allocated to previously unallocated targets (for observations in
2004 and 2005 only).  This was done by
setting any object with a good quality ID to have priority=1 (the
lowest).  Then the {\small CONFIGURE} program was re-run, but with
the fibre allocations to objects which still needed further
observation locked in place.  This was particularly
useful in quickly removing narrow emission line galaxies (NELGs) which
are often clearly identifiable in only 2 hours of observation.
Information on the observed 2SLAQ fields is presented in Table
\ref{tab:fields1}.   This lists the number of objects observed in each
field, the number of QSOs and the fraction of good quality IDs.  These
quantities are listed for the primary fibre allocation; i.e. the
sources targeted in the first night's observation of each field.  All
of these targets will have the full exposure time or have high quality
IDs in shorter exposure times.  We also list the numbers and
completeness for all the targets in each field, including those only
observed on the second (or subsequent) night.  In principle, these
could have lower completeness, as they have had shorter than average
total exposure times.

\subsection{Data reduction}

The data from the 2dF spectrographs were reduced using the {\small 2DFDR} data
reduction software \cite{2dfdr}.  Observations of a typical field
contain a fibre flat field, a calibration arc, $4\times1800$s
object frames and a final calibration arc.  The fibre flat
field frame is used to trace the positions of the fibres across the
CCD and determine the spatial profiles of the spectra for optimal
extraction, as well as to flat field the spectra to remove fibre-to-fibre
variations in spectral response.  For the object frames, fibre
throughput is calibrated using the flux in a number of strong night
sky lines and a median sky spectrum, scaled by the strong sky lines,
is then subtracted.    The object frames are combined using a variance
weighting and an additional weight (per frame) based on the mean flux
in each frame.  This accounts for variable seeing, cloud
cover etc.  Various modifications were made to the {\small 2DFDR}
software for the 2SLAQ project.  These include improvements to allow
combining of data for the same object taken in different
configurations and providing more robust methods of weighting frames.
Improvements were also made to the wavelength calibration and flat
fielding.  We note that the spectra are not spectrophotometrically
calibrated.

Data were reduced on the night of observation by team members present
at the telescope.  This operational approach has the advantage of
pseudo-real-time quality 
control of the data.  If the required spectroscopic completeness was
not achieved (80 per cent quality 1 identifications; see Section
\ref{sec:specid} below), the exposure time was extended.

\subsection{Spectroscopic identification}\label{sec:specid}

In most cases spectroscopic identification was also performed on the
night of observation at the telescope.  This enabled targets which had
sufficient $S/N$ for a good (quality 1) identification to be removed
from the configuration of the given field on subsequent nights; the
newly available fibres were then allocated to other targets.
Identification of QSO candidates was carried out in a two stage process.
First the automated identification software, {\small AUTOZ}, was used
to determine the redshift and type (e.g. QSO, star etc.) of the
object.  These automated identifications were checked using the
{\small 2DFEMLINES} software, which allows users to check the
identifications by eye and interactively adjust the
identification if required.  Both {\small AUTOZ} and {\small 2DFEMLINES}
were written for the 2QZ; details of the code are given by Croom et
al. (2001b) and C04.  Briefly, {\small AUTOZ} relies on a
$\chi^2$-minimization technique, comparing an observed spectrum to a 
number of (redshifted) template spectra.  Based on this fitting, the
spectra are classified into six categories:
\begin{tabbing}
1\=12345678\= \kill
\>{\bf QSO: } \> Broad ($> 1000\,$\kms) emission lines.\\
\>{\bf NELG:} \> Narrow ($< 1000\,$\kms) emission lines only.\\
\>{\bf gal: } \> Galaxy absorption features only.\\
\>{\bf star:} \> Stellar absorption features at $z=0$.\\
\>{\bf cont:} \> No emission or absorption features (High S/N).\\
\>{\bf ??:}   \> No emission or absorption features (Low S/N).
\end{tabbing}
A broad absorption line (BAL) QSO template was included,
and when verified by eye, BAL QSOs were labelled as ``QSO(BAL)'' in
the final catalogue.  Of 2SLAQ QSOs above $z=1.5$, where \civ\ is
visible in the observed spectrum, 171/4591 (3.7 per cent) are
classified as BALs.  This is a lower limit to the total BAL fraction
as we have not performed a consistent and quantitative analysis for
BALs [e.g. using the BALnicity index of Weymann et al.(1991)].  As a part
of the identification process, each spectrum is assigned a quality for the
identification and redshift as follows:
\begin{tabbing}
1\=1234567890123\= \kill
\>{\bf Quality 1:} \>  High-quality identification or redshift.\\
\>{\bf Quality 2:} \>  Poor-quality identification or redshift.\\
\>{\bf Quality 3:} \>  No identification or redshift assignment.\\
\end{tabbing}
The quality flag was determined independently for the identification
and redshift of an object.  For example, a quality 1 QSO
identification could have a quality 1 or 2 redshift.  A quality 1
identification is assigned if multiple spectral features are seen.
QSOs with only a strong broad \mgii\ emission line are also given a
quality 1 identification.  A quality 2 identification is given if
the is only a single spectral feature, or features of only marginal
significance.  The reliability of the different qualities is assessed
below.   

\section{The 2SLAQ QSO catalogue}\label{sec:cat}

\begin{table*}
\centering
\caption{Format for the 2SLAQ QSO catalogue.  The format entries are based
on the standard {\small FORTRAN} format descriptors.}
\label{tab:catformat}
\begin{tabular}{@{}lrl@{}}
\hline
Field&Format&Description\\
\hline
Name       &          a19 & IAU format object name\\
Priority   &           i1 & Configuration priority for 2dF\\
RA         &        f10.6 & RA J2000 in decimal degrees\\
Dec        &        f10.6 & Dec J2000 in decimal degrees\\
SDSSrun    &           i4 & SDSS run number\\
SDSSrerun  &           i2 & SDSS rerun number\\
SDSScamcol &           i1 & SDSS camera column\\
SDSSfield  &           i3 & SDSS field\\
SDSSid     &           i4 & SDSS object id within a field\\
SDSSrow    &         f8.3 & SDSS CCD Y position (pixels)\\
SDSScol    &         f8.3 & SDSS CCD X position (pixels)\\
um         &         f6.3 & SDSS PSF magnitude in $u$ band (no extinction correction)\\
gm         &         f6.3 & SDSS PSF magnitude in $g$ band (no extinction correction)\\
rm         &         f6.3 & SDSS PSF magnitude in $r$ band (no extinction correction)\\
im         &         f6.3 & SDSS PSF magnitude in $i$ band (no extinction correction)\\
zm         &         f6.3 & SDSS PSF magnitude in $z$ band (no extinction correction)\\
umerr      &         f5.3 & SDSS PSF magnitude error in $u$ band\\
gmerr      &         f5.3 & SDSS PSF magnitude error in $g$ band\\
rmerr      &         f5.3 & SDSS PSF magnitude error in $r$ band\\
imerr      &         f5.3 & SDSS PSF magnitude error in $i$ band\\
zmerr      &         f5.3 & SDSS PSF magnitude error in $z$ band\\
umred      &         f5.3 & Extinction in $u$ band (mags)\\
gmred      &         f5.3 & Extinction in $g$ band (mags)\\
rmred      &         f5.3 & Extinction in $r$ band (mags)\\
imred      &         f5.3 & Extinction in $i$ band (mags)\\ 
zmred      &         f5.3 & Extinction in $z$ band (mags)\\ 
sg         &         f8.5 & SDSS Bayesian star-galaxy classification probability\\
morph      &           i1 & SDSS Object image morphology classification 3=Galaxy, 6=Star\\
zemsdss    &         f7.4 & SDSS spectroscopic redshift\\
typesdss   &           a7 & SDSS spectroscopic identification type\\
qualsdss   &         f6.4 & SDSS spectroscopic quality\\ 
bj         &         f5.2 & 2QZ $\bj$ magnitude (Smith et al. 2005)\\
zem2df     &         f7.4 & 2QZ spectroscopic redshift (C04)\\
type2df    &           a8 & 2QZ spectroscopic identification type (C04)\\
qual2df    &           i2 & 2QZ spectroscopic identification/redshift quality (C04)\\
name2df    &          a19 & 2QZ IAU format name\\
z          &         f7.4 & 2SLAQ spectroscopic redshift\\
qual       &           i2 & 2SLAQ spectroscopic quality (ID quality $\times$ 10 + redshift quality)\\
ID         &          a10 & 2SLAQ spectroscopic identification (i.e. QSO, NELG, star etc.)\\
date       &           i6 & 2SLAQ spectroscopic observation date (YYMMDD)\\
fld        &           a3 & 2SLAQ spectroscopic field\\
fib        &           i3 & 2SLAQ spectroscopic fibre number\\   
S/N        &         f7.2 & 2SLAQ spectroscopic signal-to-noise in a 4000--5000{\AA} band\\
dmag       &         f6.2 & 2SLAQ (gm mag) - (fibre mag) relative to mean z.p. in field\\
RASS       &         f7.4 & RASS X-ray flux, ($\times10^{-13}\,$erg$\,$s$^{-1}$cm$^{-2}$)\\
FIRST      &         f6.1 & FIRST 1.4GHz Radio flux (mJy)\\
FIRSText   &           i1 & FIRST morphology; 0=no detection, 1=unresolved, 2=extended, 3=multiple\\
comment    &          a20 & 2SLAQ comment on spectrum\\
\hline
\end{tabular}
\end{table*}

\begin{table*}
\centering
\caption{Format for the 2SLAQ QSO repeated objects catalogue.  This
  lists the observational details and identifications for each object
  that was observed multiple times.  One observation is given per
  line and they are given in order of the date of observation.  The
  format entries are based on the standard {\small FORTRAN}
  format descriptors.}
\label{tab:repcatformat}
\begin{tabular}{@{}lrl@{}}
\hline
Field&Format&Description\\
\hline
Name       &          a19 & IAU format object name\\
z          &         f7.4 & 2SLAQ spectroscopic redshift\\
qual       &           i2 & 2SLAQ spectroscopic quality (ID quality $\times$ 10 + redshift quality)\\
ID         &          a10 & 2SLAQ spectroscopic identification (i.e. QSO, NELG, star etc.)\\
date       &           i6 & 2SLAQ spectroscopic observation date (YYMMDD)\\
fld        &           a3 & 2SLAQ spectroscopic field\\
fib        &           i3 & 2SLAQ spectroscopic fibre number\\   
S/N        &         f7.2 & 2SLAQ spectroscopic signal-to-noise in a 4000--5000{\AA} band\\
dmag       &         f6.2 & 2SLAQ (gm mag) - (fibre mag) relative to mean z.p. in field\\
obs        &           i1 & Number of observation for this object.\\
comment    &          a20 & 2SLAQ comment on spectrum\\
\hline
\end{tabular}
\end{table*}

In this section we discuss the 2SLAQ QSO catalogue.  The format of the catalogue is given
in Table \ref{tab:catformat}.  It is available in electronic form from
{\tt http://www.2slaq.info/}.  A second table which contains details
of multiply observed sources is also available.  The format for this
list is given by Table \ref{tab:repcatformat}.  The catalogue
includes SDSS photometry
(PSF magnitudes) and star-galaxy classification.  Note that some of
the SDSS values, such as SDSSid number, are specific to DR1, and can
change in subsequent data releases.  Where available, we
also include SDSS and 2QZ spectroscopic identifications for 2SLAQ
sources.  For the 2SLAQ spectroscopy we list the best measured
redshift, the object identification (e.g. QSO, NELG etc) and the
redshift/ID quality.  For objects with repeated observations, the
catalogue lists the parameters for the best spectrum, which is
selected based on redshift/ID quality and S/N. As part of the data
release we also provide the parameters for all other repeat
observations.  We include a number of observational details such as
date, field, fibre number and S/N (averaged in the $4000$ to
$5000$\AA\ band).  Objects which were only configured in a field on
the second or subsequent nights have been given fibre numbers greater
than 200.  The dmag entry is the difference between the observed fibre magnitude
(at $4000$ to $5000$\AA) and the SDSS PSF magnitude in the $g$-band.
This is zero-pointed to the mean difference in each field, and so
gives an estimate of which objects were brighter or fainter than their
SDSS photometry would predict.  We matched to the ROSAT All Sky Survey
(RASS; Voges et al. 1999; Voges et
al. 2000) with a maximum matching radius of 30 arcsec.  Non-matches
are indicated by zero flux in the RASS column.   We also
searched for matches to the  FIRST radio survey \cite{first}.  Given
that radio morphologies can often be complex and extended, we first
made a list of all 2SLAQ sources which had a radio match within 1
arcmin.  Each of these matches was then examined by eye to
determine whether it was a true match.  If multiple components were
present, the flux from these was summed.  The FIRSText flag is then
set based on the morphology, either unresolved (1), single extended
source (2) or multiple source (3).  The flag is set to 0 for a
non-detection.  The final entry in  the catalogue is reserved for any
comments that are made on the 2SLAQ spectrum in the process of manual
checking of the data. 

As well as the catalogue, we also make public all the spectra of
objects targetted as part of 2SLAQ observations.  These are
available as individual FITS format spectra and include repeated
observations.  A small fraction of spectra have bad 'fringing' caused
by a damaged fibre, showing up as a strong oscillation as a
function of wavelength.  These are noted as such in the comments field
of the catalogue.  Access to the spectra is via the web site {\tt
  http://www.2slaq.info/}.

We now discuss the catalogue composition and the robustness of the
identifications and redshifts.

\subsection{Catalogue composition}

\begin{figure}
\centering
\centerline{\psfig{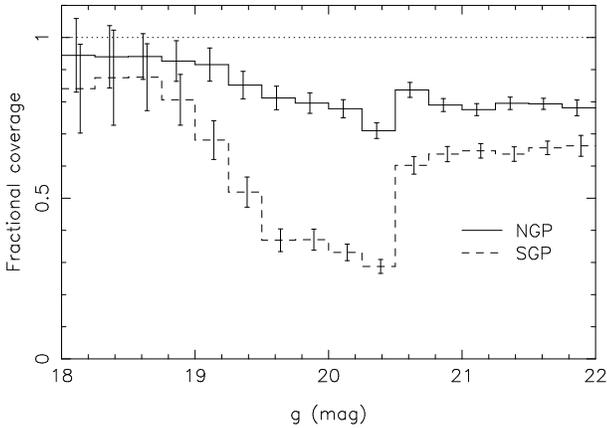}}
\caption{The fractional spectroscopic coverage as a function of
  $g$-band magnitude (extinction corrected) for the NGP (solid
  histogram) and SGP (dashed histogram) strips of the 2SLAQ QSO
  sample.  Errorbars are Poissonian.  In this plot we include objects
  which have previously been identified by the 2QZ and SDSS surveys.
  The lower coverage in the SGP
  strip is because a smaller fraction of overlapping fields were
  observed in this strip.  Also this region does not include 2QZ
  objects.  The increase towards bright magnitudes is due to the
  inclusion of SDSS identifications and the step at $g=20.5$ is due to
  our prioritization of sources fainter than this limit.}
\label{fig:covcomp}
\end{figure}

\begin{figure}
\centering
\centerline{\psfig{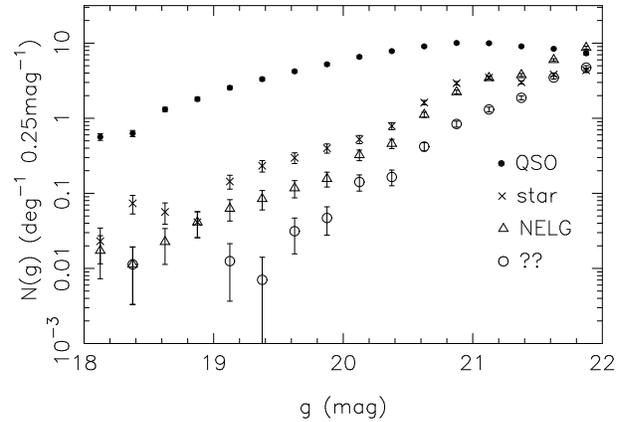}}
\caption{The number counts of sources of different types from the
  2SLAQ survey (including identifications from SDSS and 2QZ).  We show
  the QSO (filled circles), star (crosses), NELG (triangles) and
  unidentified source (open circles) number counts as a function
  of $g$-band magnitude.  Errorbars are Poissonian.  We correct the
  number counts only for the effective area covered as a function of
  magnitude (i.e. using the results in Fig.~\ref{fig:covcomp}).  The
  QSOs are the largest population at all but the faintest bin which
  has lower completeness and higher contamination (from increasing
  photometric errors and larger numbers of faint galaxies).}
\label{fig:nmobj}
\end{figure}

\begin{figure}
\centering
\centerline{\psfig{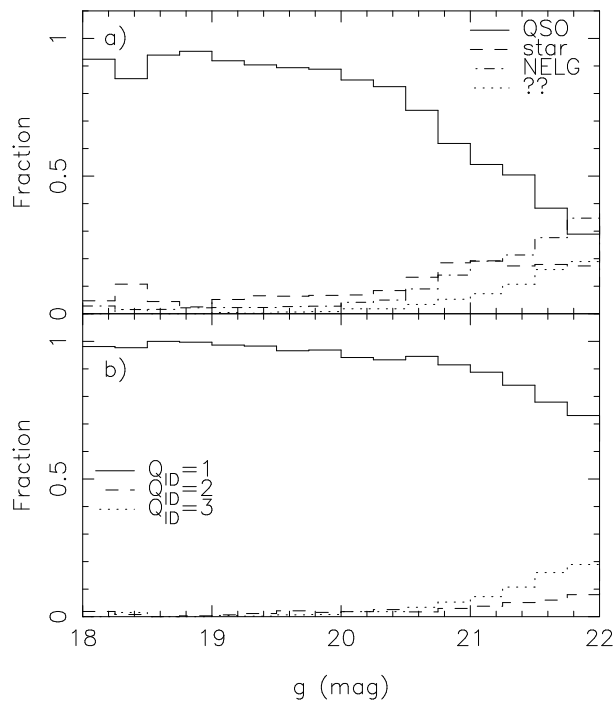}}
\caption{ a) The fraction of spectroscopically observed objects within
  2SLAQ that are QSOs (solid line) stars (dashed line), NELGs/gals
  (dot-dashed line) and unidentified (dotted line) as a function of
  $g$. b) The fraction of spectroscopic identifications with quality 1
  (solid line), 2 (dashed line) and 3 (dotted line) as a function of
  $g$.}
\label{fig:objfrac}
\end{figure}

The 2SLAQ survey regions cover 159.5~\sqdeg\ in the NGP strip and
234.1~\sqdeg\ in the SGP strip.  However, the effective area is
reduced as not all of this area was observed spectroscopically with
2dF.  This is particularly the case in the SGP.  The area covered by
2dF spectroscopy is 127.7~\sqdeg\ in the NGP and only 64.2~\sqdeg\ in
the SGP.  Spectra
were not obtained for all the sources within the observed areas
due to the high surface density of sources.  The mean target densities
were 133 and 142~\isqdeg\ for the NGP and SGP regions, respectively.
The SGP density is higher due to increased stellar contamination
(lower Galactic latitude).  As
only $\sim170$ fibres were available for QSO candidates in each field,
not all objects could be targeted (even allowing for substantial
overlap of the 2dF field centres).  

The fraction for which we did obtain spectra was also a function of
magnitude, for number of reasons.  First, we
include spectroscopic identifications from the main SDSS QSO survey
which is limited to an extinction corrected $i<19.1$ (equivalent to
$g\simlt19.3$) and the 2QZ sample limited to $\bj<20.85$ (equivalent to
$g\simlt20.85$, not extinction corrected).  The resulting coverage
completeness is shown in
Fig.~\ref{fig:covcomp} as a function of magnitude.  The NGP strip has
reasonably uniform coverage, which is never below 70 per cent, while the SGP strip
(without any 2QZ spectra and more incomplete spectroscopic coverage)
varies much more, reaching $\sim30$ per cent at worst.  The
visible step in the fractional coverage at $g=20.5$ is due to our
prioritization of objects fainter than this limit (see Section
\ref{sec:targpri}).  The numbers of 
objects in the survey regions targeted by 2dF are listed in
Table~\ref{tab:nobj} as a function of priority, for NGP and SGP strips
separately.  At bright magnitudes ($g<20.5$)
approximately equal number of spectra are contributed from previous
surveys (SDSS and 2QZ) and 2SLAQ, while at fainter magnitudes 96 per
cent of the spectroscopic observations are new.

\begin{table}
\begin{center}
\caption{The number of objects within 2SLAQ 2dF fields ($N_{\rm obj}$)
  as a function of priority.  We shown this separately for the NGP and
  SGP strips.  We also list the number of objects in the same regions
  with spectroscopy from SDSS (DR4), 2QZ and 2SLAQ as well as the
  total number of objects with spectroscopic observations ($N_{\rm
  obs}$).  Because some objects were observed by more
  than one of SDSS, 2QZ and 2SLAQ, $N_{\rm obs}$ is not equal to the
  sum of the other columns.  Also, we do not include the 19 objects
  that are outside our formal 2SLAQ limits (see Table
  \ref{tab:outside}).  $F_{\rm obs}$ is the fraction of objects
  observed, i.e. $N_{\rm obs}/N_{\rm obj}$. $A_{\rm obs}$ is the
  effective area for each sample i.e. (surveyed area)$\times F_{\rm
  obs}$.}
\setlength{\tabcolsep}{4pt}
\begin{tabular}{cccccccc}
\hline
Pri. & $N_{\rm obj}$ & $N_{\rm SDSS}$ & $N_{\rm 2QZ}$ & $N_{\rm
  2SLAQ}$ & $N_{\rm obs}$ &  $F_{\rm obs}$ & $A_{\rm obs}$\\ 
& & & & & & &\sqdeg\\
\hline
3-NGP &   4795 &  1054 &  2351 &  2015 &  3893 & 0.812 & 102.60\\
4-NGP &    795 &    18 &     1 &   459 &   474 & 0.576 &  75.34\\
5-NGP &   4567 &     0 &    62 &  3321 &  3338 & 0.731 &  92.36\\
6-NGP &   6908 &     6 &   501 &  5733 &  6042 & 0.875 & 110.53\\
\hline
3-SGP &   2351 &   553 &     0 &   561 &  1051 & 0.447 &  28.38\\
4-SGP &     23 &     5 &     0 &     7 &    12 & 0.522 &  33.12\\
5-SGP &   2884 &     0 &     0 &  1584 &  1584 & 0.549 &  34.86\\
6-SGP &   3413 &    17 &     0 &  2627 &  2642 & 0.774 &  49.13\\
\hline
\label{tab:nobj}
\end{tabular}
\end{center}
\end{table}

\begin{figure}
\centering
\centerline{\psfig{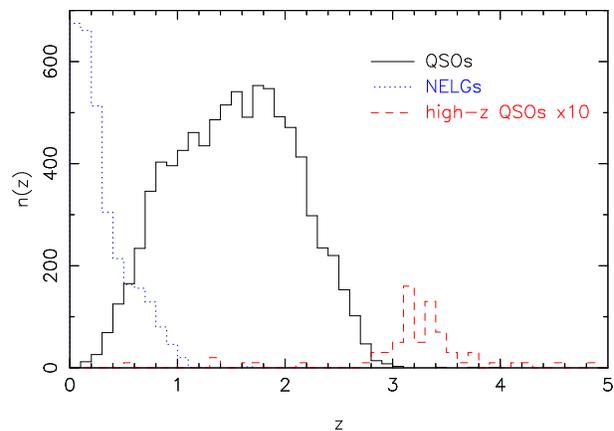}}
\caption{The redshift distribution of 2SLAQ observed sources.  We plot
  the distribution of low redshift QSOs (solid black line), NELGs
  (dotted blue line) and high redshift QSOs (dashed red line).  The
  distribution of high redshift QSOs, which are those selected via
  the high redshift colour cuts (Section \ref{sec:highzsel}), has been
  multiplied by a factor of 10.}
\label{fig:nz}
\end{figure}

The survey composition also varies as a function of magnitude; this
is shown in Fig. \ref{fig:nmobj}.  We have corrected these number
counts for incompleteness in the spectroscopic coverage as a function
of magnitude only (i.e. Fig.~\ref{fig:covcomp}), and have not included
incompleteness from colour selection or spectroscopy.  At bright
magnitudes ($g<21$) the QSOs dominate the sample, the stars are the
next largest population, followed by NELGs (we've also included here
the small number of absorption line galaxies, classified as 'gal') and
lastly, unidentified objects.  At $g>21$ QSOs are still the largest
single population, but the other populations become more significant.
This is because of the intrinsic flattening of the QSO number counts
(e.g. see R05), reduced completeness for QSOs
from increased photometric errors and increased contamination from
the large number of faint galaxies (and increased photometric
errors).  In fact, in the faintest bin the number of NELGs exceeds the
number of QSOs.  This is more clearly seen in Fig.~\ref{fig:objfrac}a
which shows the relative fraction of objects of each type.  Our
spectroscopic incompleteness (i.e. the
fraction of objects with no spectroscopic identification) increases
towards the faint limit of the sample.  Fig.~\ref{fig:objfrac}b shows
the fraction of identification qualities as a function of $g$.  The
fraction of quality 1 identifications declines to 73 per cent at the
faint limit of the survey. 
Of the remaining objects, 1/3 have quality 2 identifications and 2/3
have quality 3 identifications (the poorest).  For most analyses in
this paper, only quality 1 identifications are used.

\begin{figure*}
\centering
\centerline{\psfig{file=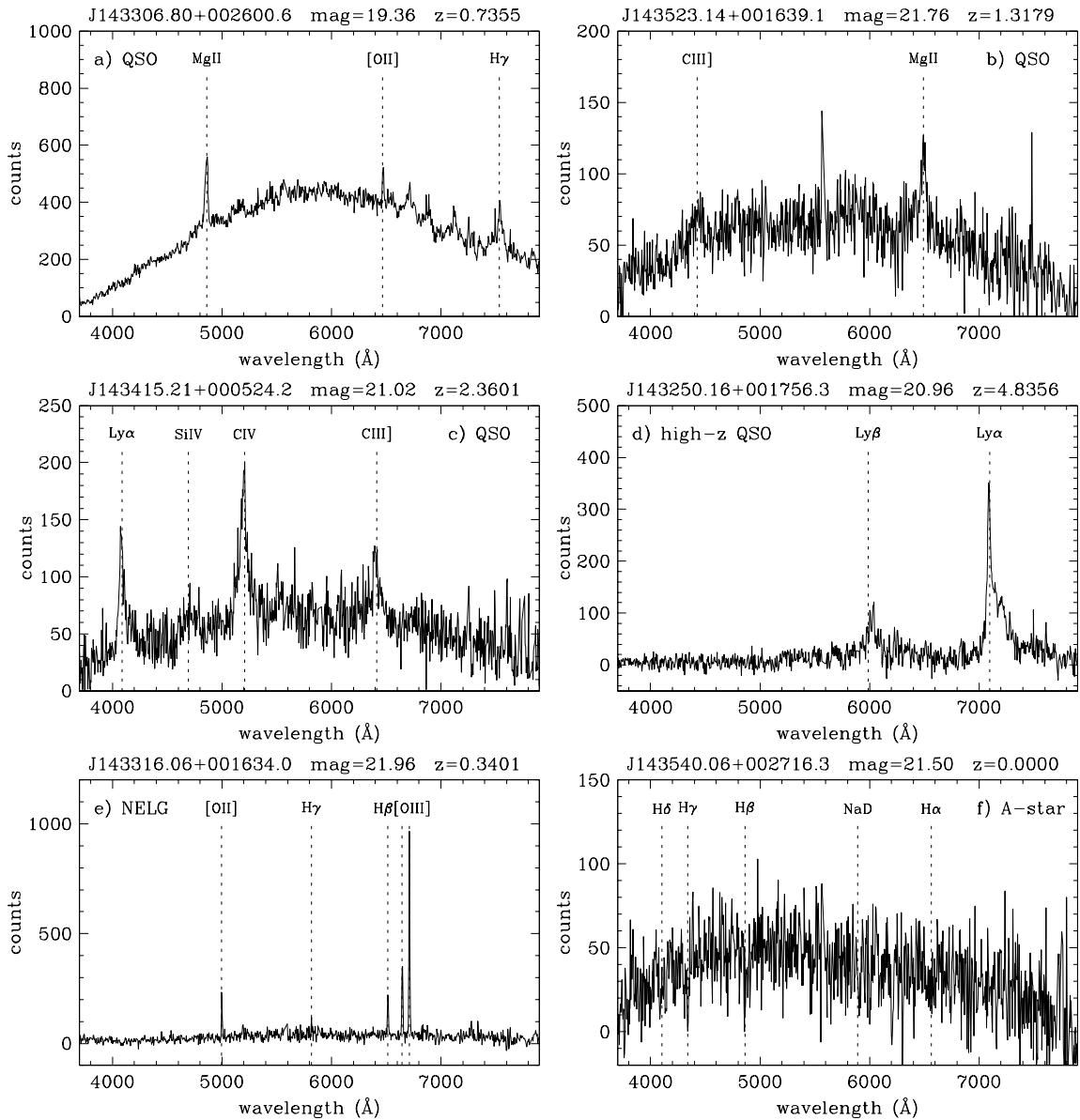,width=18cm}}
\caption{Example 2dF spectra from the 2SLAQ survey taken from
  observations of the e01 field in April 2005. a), b) and c) show QSOs
  from the main low-redshift sample, d) shows the highest redshift
  QSO from the high-redshift sample, e) is a typical NELG and f) shows
  a Galactic star.  The magnitudes displayed are PSF $g$-band
  magnitudes (not extinction corrected), apart from d) in which we
  give the PSF $i$-band magnitude.  The spectra are not
  spectrophotometrically calibrated, so the concave spectral shape is
  due to the instrument response.}
\label{fig:spec}
\end{figure*}

The redshift distribution of the main low redshift QSO sample is shown
in Fig. \ref{fig:nz} as the solid line.  The number of QSOs is
relatively constant between $z\simeq0.8$ and $2.2$, declining towards
lower and higher redshift.  The high redshift sample (Section
\ref{sec:highzsel}) primarily samples the redshift range between
$z\simeq 2.8$ and $4.0$, with the highest redshift QSO being
J143250.16+001756.3 at $z=4.8356$.  The NELGs (including some
absorption line galaxies) are peaked at low redshift, with a tail of
objects to $z\simeq1$.  Example 2SLAQ spectra, including a range of
QSO redshifts and magnitudes as well as a NELG and a Galactic star, are
shown in Fig. \ref{fig:spec}.  

\subsection{Repeatability of identifications and redshifts}\label{sec:rep}

A critical test of the quality of the catalogue is to assess the
reliability and repeatability of our 
identifications and redshifts.  We can do this both internally, using
repeat observations, and externally using comparisons to other
catalogues.  In particular, we have 2SLAQ spectra for objects which
also have SDSS and 2QZ spectra available.  In this external check it
is worth noting that the overlap between SDSS/2QZ and 2SLAQ is only at
the bright end of the sample, where identifications are inherently
more reliable.  Secondly, the 2QZ is only nominally an external check,
as the data acquisition, reduction and analysis for 2QZ and 2SLAQ are
almost identical.

We start by assessing the relative reliability of the identifications
between surveys.  In Table \ref{tab:failures} we list the number of
objects with different quality identifications in more than one
sample, compared to the number for which that identification disagrees
between samples.  For good quality identifications ($Q_{\rm ID}=1$) in
two samples we find that 4/516 objects disagree between SDSS and 2QZ
($0.8\pm0.4$ per cent), 4/378 between SDSS and 2SLAQ ($1.1\pm0.5$ per
cent) and 15/771 between 2QZ and 2SLAQ ($2.0\pm0.5$ per cent).  By
visually examining the spectra we find that for all four SDSS-2QZ
discrepancies the SDSS identification is correct.  For the SDSS-2SLAQ
comparison we find the SDSS identification to be correct in 3 cases
and the 2SLAQ identification to be correct in 1 case.  All 4 objects
have the same redshift in both samples, and the disagreement is due to
classification as NELG or QSO.  For the 2QZ-2SLAQ comparison we found 5
cases in which the 2QZ identification was correct and 10 cases were
the 2SLAQ identification was correct.  Of these 10, eight were
low $S/N$ objects in 2QZ wrongly classified 
as stars.  If we compare lower quality identifications we
find that, as expected, the repeatability is poorer (final three
columns in Table \ref{tab:failures}).

\begin{table}
\begin{center}
\caption{The fraction of objects which have different identifications
  (i.e. QSO vs. star etc.) in the different surveys
  (SDSS/2QZ/2SLAQ).  1-1 denotes a quality 
  1 ID for both the first and second catalogue listed in the first
  column; 1-2 denotes a quality 1 and a quality 2 ID for the
  first and second catalogues listed etc.} 
\setlength{\tabcolsep}{4pt}
\begin{tabular}{ccccc}
\hline
Surveys & $Q_{\rm ID}$ &  $Q_{\rm ID}$ &  $Q_{\rm ID}$ &  $Q_{\rm ID}$\\
        &  1-1         &    2-2     & 1-2  &  2-1 \\
\hline
SDSS-2QZ   & 4/516  &  0/0 &  5/11 &   0/19\\
SDSS-2SLAQ & 4/378  &  0/0 &  0/1  &    0/8\\
2QZ-2SLAQ  & 15/771 &  3/3 &  2/9  &  36/90\\
\hline
\label{tab:failures}
\end{tabular}
\end{center}
\end{table}

\begin{table}
\begin{center}
\caption{The fraction of objects with the same identification which
  have different redshifts  
  in the different surveys (SDSS/2QZ/2SLAQ).   11-11 denotes a quality
  11 spectrum for both the first and second catalogue listed in the first
  column; 11-12 denotes a quality 11 and a quality 12 spectrum for the
  first and second catalogue listed etc.} 
\setlength{\tabcolsep}{4pt}
\begin{tabular}{cccccc}
\hline
Surveys & $Q_{\rm ID}$ & $Q$ & $Q$ & $Q$ & $Q$ \\
        & 1-1 & 11-11 & 11-12 & 12-11 & 12-12 \\

\hline
SDSS-2QZ   & 3/512  &  2/510 &  1/2  &   0/0  & 0/0 \\
SDSS-2SLAQ & 2/374  &  2/374 &  0/0  &   0/0  & 0/0 \\
2QZ-2SLAQ  & 20/756 &  10/720 & 1/5  &   9/31 & 0/0 \\
\hline
\label{tab:failuresz}
\end{tabular}
\end{center}
\end{table}

Next we check the external reliability of the redshift
estimates. We search for objects identified in more than one sample
that have the same identification (with $Q_{\rm ID}=1$) but a redshift
difference of greater than $\Delta z=0.05$ (chosen to include only
those objects with catastrophic failures in redshift). This resulted
in 3/512 for the SDSS-2QZ comparison, 2/374 for the SDSS-2SLAQ
comparison and 20/756 in the 2QZ-2SLAQ comparison.  However, for a
number of objects the redshift was flagged as uncertain (i.e. $Q_{\rm
  z}=2$).  If we limit ourselves to objects with $Q=11$ (i.e. $Q_{\rm
  ID}=1$ and $Q_{\rm z}=1$) then the fractions are 2/510, 2/374 and
10/720 respectively (see Table \ref{tab:failuresz}).  Visual
assessment of the spectra for which there were discrepant redshifts
showed that the surveys scored 2-1 (SDSS-2QZ), 2-0 (SDSS-2SLAQ) and 1-19
(2QZ-2SLAQ).  The most common error was a confusion between \mgii\ and
\civ, particularly when \ciii\ was weak.

We can similarly check the reliability of our identifications and
redshifts with internal checks using the repeat observations of 2SLAQ
sources.  A total of 3317 objects were repeated, of which  2911 had two
observations, 382 had three observations, 23 had four observations and
a single object had five observations.  These repeats were generally
made when overlapping fields were observed, and are biased towards
objects which had a poor identification in one or more observation.
However, there are 1672 objects with at least two observations which
have both have quality 11.  Of these, 1585 had identical identification
and redshift ($\Delta z<0.05$) and 87 did not match.  Thus the quality
11 objects are reliable at the 95 per cent level.  Of the 87 that do
not match, 33 have the same identification, but a different redshift
(mostly QSOs, but with some NELGs).  The remainder (54) have different
identifications, and of these 12 have the same redshift but were
identified as QSO from one spectrum and NELG in the other.  This is
usually due to a weak broad component combined with a stronger narrow
component in the \hb\ line that is not identified in one spectrum
(typically that with lower $S/N$).  

\begin{figure}
\centering
\centerline{\psfig{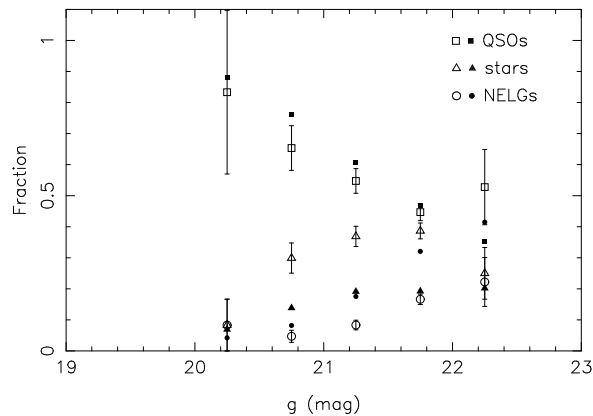}}
\caption{The fraction of good quality identifications that are QSOs,
  stars and NELGs (filled squares, triangles and circles respectively)
  compared to the fraction of poor ($Q_{\rm ID}=2$ or 3)
  identifications that are QSOs, stars and NELGs (open squares,
  triangles and circles respectively), derived from repeat
  observations.  Errors are only plotted for the open symbols.  For
  the filled symbols the errors are generally the size of the symbol
  or smaller.}
\label{fig:fractypes}
\end{figure}

Another important question to
address is the content of the unidentified objects.  At the faint
limit of the sample, this reaches 27 per cent (quality 2 and 3
identifications); see Fig. \ref{fig:objfrac}b.  A first order
assessment of the content of these low quality spectra can be made by
comparing repeats that have one high quality ($Q_{\rm ID}=1$) and one
low quality ($Q_{\rm ID}=2$ or 3) spectrum.  This enables us to assess
the fraction of QSOs, stars and NELGs that are contained within the
unidentified objects.  Fig. \ref{fig:fractypes} shows the fraction of
these repeats that have a good ID that is a QSO, star or NELG (open
symbols).  This is compared to the fractions among all the high
quality objects (i.e. not just those with repeats; filled symbols).
We see that at all magnitudes the QSO fraction in the identified and
unidentified objects is similar.  The fraction of NELGs in the unidentified
objects is significantly lower than in the whole sample, while the
fraction of stars is higher.  This is to be expected given the ease of
identifying NELGs with their strong narrow emission lines and the
difficulty of identifying stars with their relatively weak absorption
features.  This analysis does not account for all the
unidentified objects in our sample, as some are not identified even
with a second observation.  In the half magnitude bins from $g=20$ to $g=22.5$
used in Fig. \ref{fig:fractypes}, the fraction of poor spectra not
identified in a second observation is 8, 11, 22, 31 and 51 per cent
from bright to faint magnitudes.  Therefore, for all but the faintest
bin (which contains a small number of objects with high extinction
and the high redshift QSO candidates), most objects are identified
with a second spectrum.

Finally we assess the internal reliability of our redshift estimates.
There are 1672 objects with repeats that are both quality 11.  Of
these 33 (2.0 per cent) have different redshifts.  There are 156
repeated objects with both a quality 11 and 12 observation, of which
37 (24 per cent) have different ($\Delta z>0.05$) redshifts.  This
higher fraction is expected for objects with quality 2 redshift
determinations.

\begin{figure*}
\centering
\centerline{\psfig{file=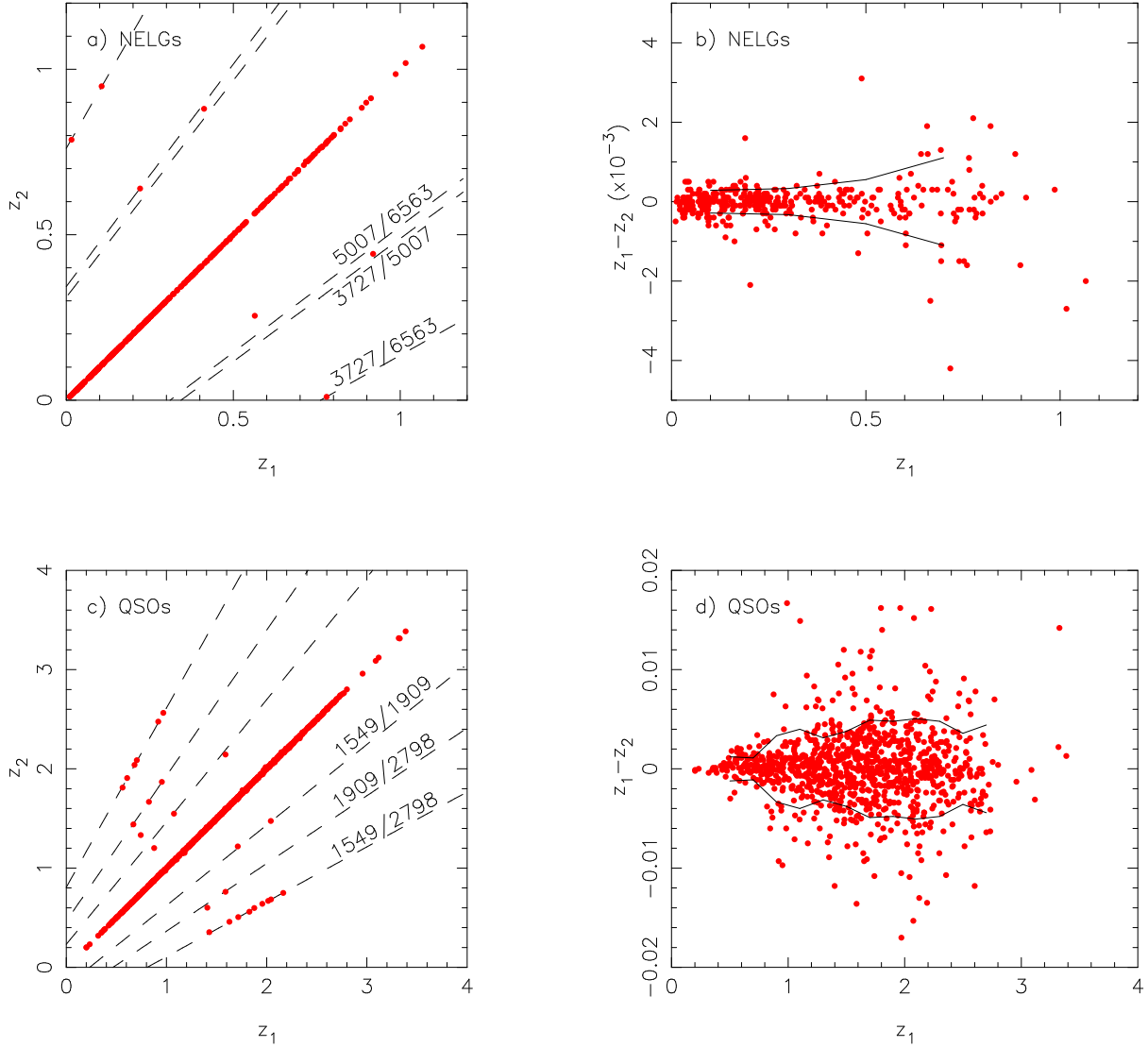,width=16cm}}
\caption{a) Comparison of redshifts from repeat observations of
  NELGs.  Most lie close to the $z_1=z_2$ line, while there are a few
  outliers.  The dashed lines show the redshift differences caused by
  confusing specific emission lines (rest wavelengths shown).  b)
  Redshift difference vs. redshift for NELGs, the solid lines show the
  RMS scatter as a function of redshift. c) and d) are the same as a)
  and b) only for QSOs rather than NELGs.  In all cases only quality
  11 observations were considered.} 
\label{fig:zdisp}
\end{figure*}

We assess the accuracy of our redshift estimates by examining
the scatter in redshifts for repeat observations (quality 11 only);
see Fig. \ref{fig:zdisp}.
We do this separately for NELGs and QSOs, as the NELGs with their
narrow lines have a much smaller dispersion than the QSOs.  Most
repeat redshifts lie close to the line 
$z_1=z_2$.  However a small number lie along lines denoting incorrect
emission line identifications (dashed lines in Fig. \ref{fig:zdisp}).
7/385 NELGs (1.8 per 
cent) with repeats have discrepant redshifts due to mis-identification
of emission lines.  26/1032 QSO repeats (2.5 per cent) have discrepant
redshifts, of which half are due to confusion between \mgii\
and \civ.  The scatter in redshift measurement after removing these
catastrophic failures is shown in Figs. \ref{fig:zdisp}b and d for
NELGs and QSOs respectively.  The solid lines show the RMS scatter in
$\Delta z=0.2$ bins (calculated for bins with greater than 10
objects).  The scatter in NELG redshifts is small,
$\sigma(z_1-z_2)=0.00044$ ($\simeq 130$\kms).  The scatter 
increases with redshift, being 0.00029 at $z=0.1$ and 0.0011 at
$z=0.7$.  A similar trend is seen for the QSOs, but with a larger mean
scatter of $\sigma(z_1-z_2)=0.0037$.  We find that
$\sigma(z_1-z_2)=0.0014(1+z)$ gives a good description of the scatter
in QSO redshifts as a function of redshift, identical to that found by
Croom et al. (2005) for the 2QZ (this is not surprising given the
identical spectrographic configuration and similar data quality).

\section{Survey completeness}\label{sec:comp}

We now discuss quantitative assessments of the completeness
of the 2SLAQ QSO sample.  In general we can separate the completeness
into four distinct types, which are a function of $g$-band magnitude,
redshift $z$ (below $z$ denotes redshift rather than $z$-band
magnitude) and celestial position ($\alpha,\delta$).

\begin{itemize}
\item {\bf Morphological completeness}, $f_{\rm m}(g,z)$.
  This describes our effectiveness at differentiating between point
  and extended sources in the SDSS imaging.  We include extended sources
  in the low redshift QSO sample.
\item {\bf Photometric completeness}, $f_{\rm p}(g,z)$.  This
  attempts to take into account any QSOs which may have fallen outside
  our colour selection limits.  
\item {\bf Coverage completeness (or coverage)}, $f_{\rm c}(\alpha,\delta,g)$.
  This is the fraction of 2SLAQ
  sources which have spectroscopic observations.
\item {\bf Spectroscopic completeness}, $\fs(\alpha,\delta,g,z)$.
  This is the fraction of objects which have spectra with quality 1
  identifications.
\end{itemize}

\subsection{Morphological completeness}

As discussed by R05, we initially included in our sample
objects that the SDSS photometric pipeline ({\small PHOTO}; Lupton et
al 2001) classify as extended.  This is because a significant number
of point sources are mis-classified as extended at the faint limit of
our sample ($\sim15$ per cent; see Fig. 5 of R05) and low redshift
QSOs can be genuinely extended.
However, our first observing runs (March and April 2003)  showed the
sample contained large numbers of NELGs, of which many were extended.
To reduce the contamination by NELGs, the final sample cuts were more
restrictive and included morphology restrictions using the Bayesian
star-galaxy classifier of Scranton et al. (2002). These are described
in detail in Section \ref{sec:sel}.  A total of 2144 objects were
observed with the preliminary colour selection, of which 1021, 590,
283 and 250 were QSOs, NELGs, stars and ?? identifications,
respectively.  Of these sources, 284 were subsequently rejected from
the final catalogue on the basis of morphology, of which 262 were
NELGs, 9 QSOs, 7 stars and 6 unclassifiable.  This morphological cut removed
44 per cent of the NELG contamination while only losing 0.9 per cent
of the QSOs.  The QSOs rejected by this cut are at low redshift,
between $z=0.12$ and $0.84$, and are distributed uniformly within this
interval.  Thus at low redshift we do lose some
QSOs due to the extended nature of their hosts.  The 7 stars
rejected (all with $g>20.7$) suggest that at the faint limit the Bayesian
star-galaxy classifier is not perfect, so that a small number of QSOs
would be lost from the sample even though their true morphology was
point-like.  Hence, the accuracy of the Bayesian
star-galaxy classifier, together with our conservative cuts, means
that the rejection of low redshift QSOs is minimal, and we will
generally not correct for it in our analysis below.

As we incorporate 2QZ redshifts into the 2SLAQ catalogue, 
another issue to consider is that the 2QZ selection only included
point-sources from APM scans of UKST photographic plates.  C04
and Smith et al. (2005) discuss the morphological selection
of the 2QZ in detail.  There are two types of morphological
incompleteness.  The first is due to objects which are true point sources
but which the APM software has classified at extended.  From
comparisons to SDSS imaging data Smith et al. (2005) show this to be a
weak function of magnitude, rising from 6.4 per cent at the bright end
and 8.9 per cent at the faint end.  Secondly, there are objects which are
truly extended (and classified as such from UKST plates), and are
therefore missed by the 2QZ selection.  C04 argue that
this should only be a significant effect at $z<0.4$ in the 2QZ given
that the typical size of stellar images on UKST plates is 2--3
arcsec.  In principle, the morphological bias of 2QZ objects could
impact our 2SLAQ catalogue.  This is because not all 2SLAQ selected
targets have been observed spectroscopically, and those with 2QZ
redshifts will preferentially be point sources.  The vast majority of
2SLAQ sources with 2QZ spectra have $g=19.0-21.0$; in this interval
the fraction of all QSOs
which are classified as extended by SDSS (SDSS TYPE = 3) was $10.3\pm0.4$ per
cent.  In contrast, $7.6\pm0.6$ per cent of 2SLAQ QSOs which have
2QZ spectra are classified as extended.  Therefore, $2.7\pm0.7$ per
cent of QSOs may be missed if only targeted with 2QZ.  Given
the coverage completeness in the range $\sim70-80$ per cent for the
NGP (Fig. \ref{fig:covcomp}), the actual morphological bias introduced
by the 2QZ selection will only be at the 0.5--0.8 per cent level at
worst.  We therefore do not account for this insignificant bias in our
analysis below.

\subsection{Photometric completeness}\label{sec:photcomp}

In order to determine the completeness of our sample, we construct a
set of model QSO colours.  In doing this we aim to trace as accurately
as possible the underlying distribution, including the evolution of
colour with redshift and a detailed consideration of the effects of
host galaxies.  These model colours are then passed through our
selection algorithm to estimate the fraction of objects selected.  We
use a modified version of the technique described by R05 (and Fan
1999), but unlike them, we also incorporate the impact of host
galaxies. 

\subsubsection{Simulating QSO colours}\label{sec:qsosim}

We start by generating a set of QSO-only spectral energy distributions
(SEDs) (i.e. not including host galaxy contributions) which are
well matched to the colours of bright ($i<18$) SDSS QSOs taken from
the DR3 catalogue \cite{sdssqso3}.  These are similar to those
generated by various other authors (e.g. Fan 1999; Richards 2006).
The underlying continuum is assumed to be a power law in frequency,
$\nu$, of the form $f_\nu\propto\nu^{\alpha_\nu}$ (equivalent
to a power law in wavelength of $f_\lambda\propto\lambda^{-[\alpha_\nu+2]}$).
The power law index, $\alpha_\nu$ is normally distributed with a
mean $\alpha_\nu=-0.3$ and a standard deviation of 0.3.  A mean of
$\alpha_\nu=-0.3$ is slightly bluer than that assumed by some other
authors (e.g. Richards et al. (2006) used $<\alpha_\nu>=-0.5$).
We find that the bluer $\alpha_\nu=-0.3$, provides a better match to
the observed colours of bright 
QSOs.  This may be because the measured redder slopes already include
some contribution from their host galaxy.

We then include an emission line component
using the SDSS QSO composite \cite{sdsscomp}.  We divided the composite
spectrum by a fit to the continuum at $\lambda<5000$\AA\ then subtract
a second power law red-ward of this limit.  The composite is seen to
have a redder spectrum at $\lambda>5000$\AA; probably due to contamination of low
redshift QSOs by a host galaxy contribution, which we wish to remove
before creating our emission line spectrum.  Hence we subtract the
continuum red-ward of this limit rather than dividing by it.  All
emission lines are assumed to have the same {\it relative} equivalent
width (EW), but the emission line spectrum is scaled by a factor that
has a log-normal distribution with a mean of 1 and $\sigma=0.48$,
consistent with the measurements made of the EW distribution of 2QZ
QSOs by Londish (2003).  We also make a correction to the flux around the
Lyman-$\alpha$ line to account for absorption already present in the
composite spectrum.  To match the $u-g$ colours of QSOs at
$z\sim1.5-2.2$ we boost the flux at $\lambda=1180-1290$\AA\ by a
factor of 1.2.  We add a Balmer continuum (BC) component to
the QSO spectrum of the form given by Eq. 6 in Grandi (1982).  We
use a temperature of 12000K and the relative normalization of the BC
component that is 0.05 of the underlying continuum at 3000\AA, which
we find matches the observed colours of bright QSOs.  This
is somewhat lower than the 0.1 fraction used by Maddox \& Hewett
(2006) in their generation of simulated QSO spectra.  We suspect that
this difference is due to 
the fact that the emission line spectrum we use effectively
contains some fraction of the BC component as well.

\begin{table}
\begin{center}
\caption{Parameters of the assumed distribution of Lyman-$\alpha$
  forest, Lyman-limit and damped absorbers.  The evolution of the
  absorbers is modelled by a power law of the form
  $N(z)=N_0(1+z)^\gamma$ and the H~{\small I} column distribution is
  proportional to $N_{\rm H}^{-\beta}$. $b$ is the Doppler width of
  the Voigt profile.  Note that these parameters are the same
  as those given by Fan (1999) with the exception of $N_0$ for the
  Lyman-$\alpha$ forest (Fan et al. used $N_0=50$).}
\setlength{\tabcolsep}{4pt}
\begin{tabular}{cccccc}
\hline
Absorption type & $\log(N_{\rm H})$ & $N_0$ & $\gamma$ & $\beta$ & $b$\\
& (cm$^{-2}$)      & & & & (\kms) \\
\hline
Lyman-$\alpha$ forest & 13.0--17.3 & 20.0 & 2.3 & 1.41 & 30\\
Lyman-limit           & 17.3--20.5 & 0.27 & 1.55 & 1.25 & 70\\
Damped Lyman-$\alpha$ & 20.5--22.0 & 0.04 & 1.3 & 1.48 & 70\\

\hline
\label{tab:lyman}
\end{tabular}
\end{center}
\end{table}

\begin{figure*}
\centering
\centerline{\psfig{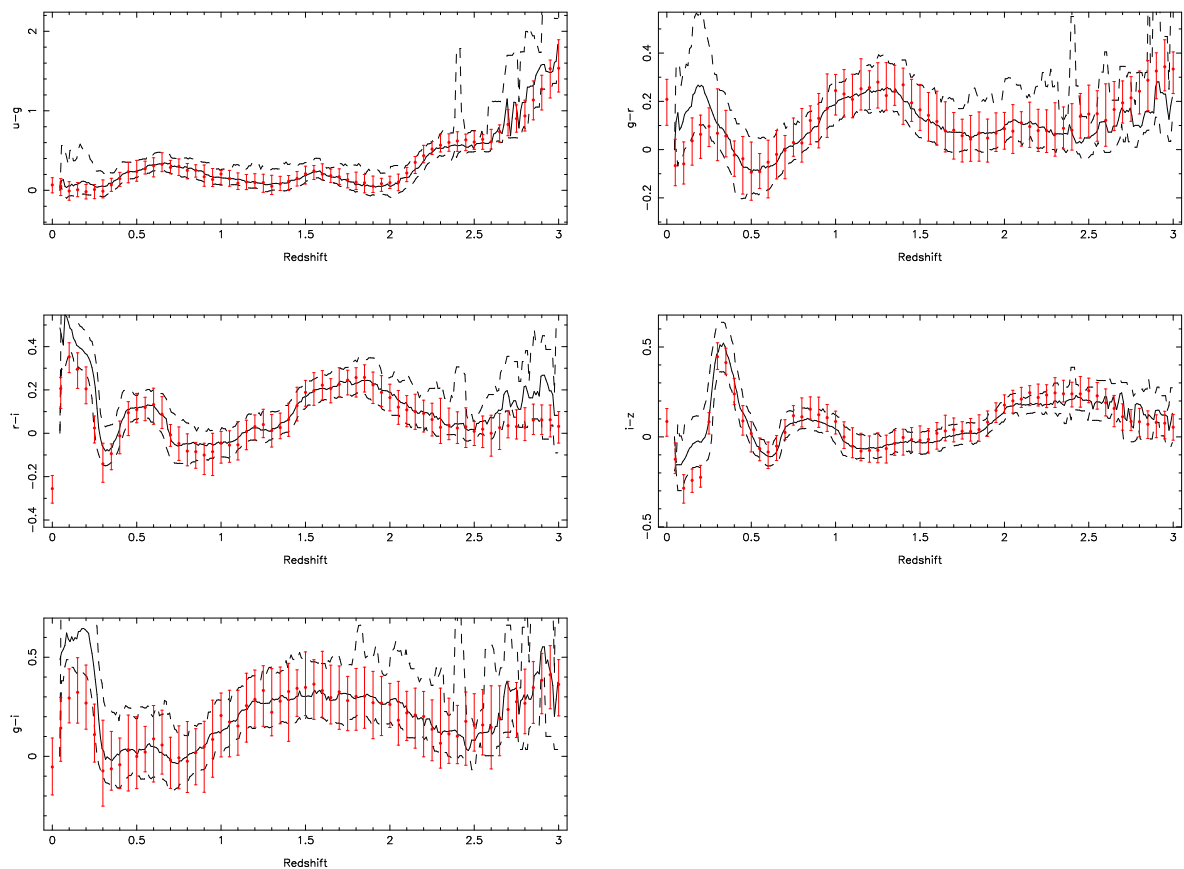}}
\caption{The median simulated QSO colours not including host galaxy
  contributions (filled circles) compared to QSOs with $i<18$ from the
  SDSS DR3 QSO catalogue (Schneider et al 2005; solid line).  The
  error bars on the points show the 68 
  percentile range for the simulated colours (equivalent to $1\sigma$
  Gaussian errors) and can be compared to the same range shown for the
  SDSS DR3 QSOs (dashed lines).  The simulated colours match the
  observed colours well at redshift $z>0.25$.  Below this redshift,
  even the relatively bright QSOs from DR3 are affected by some
  host galaxy contamination.}
\label{fig:colz}
\end{figure*}

Absorption blue-wards of Lyman-$\alpha$ is added to the spectrum
following the recipe of Fan (1999) with some minor modifications.  We
calculate the contributions to the opacity for the first 10
transitions and use the accurate approximation of Tepper Garc\'{i}a (2006) to
calculate Voigt profiles (although note that they  have an
error in the equation in footnote 4 of their paper, and $Q\equiv1.5/x^2$
rather than $Q\equiv1.5x^2$).  For each absorber the Lyman limit
absorption was calculated using the prescription of Kennefick et
al. (1995).  The forest, Lyman-limit and damped absorbers are each
distributed according to the values listed
in Table \ref{tab:lyman}, which follows closely the values listed by
Fan (1999) except for the value of $N_0$ for the Lyman-$\alpha$
forest.  We find a better match to the QSO colours with a lower value
than Fan (1999).

Next, asinh magnitudes in the SDSS bands are calculated from the
QSO SEDs, including the small corrections to transform from true AB to
the SDSS system ($u\ab = u\sdss + 0.04$ and $z\ab=z\sdss+0.02$)
\cite{aaa04}.

Finally, we add a Gaussian random error with a $\sigma$ drawn from the
SDSS PSF magnitude errors at the simulated magnitudes.  This error was
combined in quadrature with the
uncertainties in the photometric calibration (0.03 in $u$ and $z$, and
0.02 in $g$, $r$ and $i$).  The colour redshift relations for these
(QSO only) simulations are shown in Fig. \ref{fig:colz}.  In this plot
we show the four usual colours: $u-g$, $g-r$, $r-i$ and $i-z$ as a
function of redshift.  We also show $g-i$ which, although not
independent of the first four colours, is one of the primary colours used
in 2SLAQ selection.  We compare these relations to
those derived from SDSS QSOs from the DR3 QSO catalogue
\cite{sdssqso3} with $i<18$.  This brighter magnitude was chosen to
reduce the effect of host galaxy contamination on the observed colours
at low redshift.  The median simulation
colours accurately track the observed medians over almost the entire
range sampled.  At $z<0.25$ the simulated colours are
consistently bluer than the observed colours by up to $\sim0.2$ mag.
This we ascribe to the effect of host galaxy contamination in these
low redshift sources.  At $z>2.5$ there is some evidence that the
simulated $r$-band magnitudes are $\sim0.1-0.15$ mag too bright (seen
in the $g-r$ and $r-i$ vs redshift plots).  This could plausibly be
due to the models having too much \civ\ flux at this redshift.
Fig. \ref{fig:colz} also shows the 68 percentile range for the
simulations (errorbars on points) and the data (dashed lines).  These
are also in excellent agreement for all but the lowest redshift
intervals ($z<0.25$).  In summary, over the redshift range for which
the 2SLAQ survey has high completeness ($0.4<z<2.6$; see below), the
model QSO colours are an excellent match to the observed colours of
bright QSOs.

\subsubsection{Simulating host galaxy colours}\label{sec:hostgals}

Now that we have a reliable method for simulating QSO spectra, we must
consider the impact of the host galaxy on the final observed
colours of 2SLAQ objects.  The 2SLAQ selection is based on SDSS PSF
photometry, so host galaxy contributions are to some extent minimized,
but at the faint flux limits we reach these PSF magnitudes still
contain significant host galaxy contributions (e.g. Schneider et
al. 2003).  We start by
considering the observed relation between $L\qso$ and $L\gal$ from the low
redshift host galaxy analysis of Schade, Boyle \& Letawsky (2000).
Taking this data and fitting a relation
\begin{equation}
M\gal=A+BM\qso
\label{eq:galqsorel}
\end{equation}
to all objects with point source detections in the $B$-band brighter
  than $M_{\rm B}(AB)=-16$ we obtain
$A=-17.1\pm1.1$ and $B=0.21\pm0.05$ with a scatter, $\sigma\qg$=0.7 in
  $M\gal$.  If we instead assume no correlation between $M\gal$ and
$M\qso$, we find the mean $M\gal(AB)$ is
  $-21.3$ with an rms scatter of 0.8.  Within the luminosity range that 
  Schade et al.\ probe, the $M\gal-M\qso$ correlation is significant,
  but if brighter AGN are added to the sample, the
  relation appears to flatten (see Fig. 13c of Schade et al.).  With
  this in mind, below we will investigate whether we can constrain the
  slope of the relation by comparing our simulations and 2SLAQ colours.

Our approach is to constrain as many parameters of the host galaxy SED
as possible from independent observations
and then use the colour distribution of the 2SLAQ QSOs to adjust the
other parameters.  In particular, we need to ensure that when we apply
the colour selection criteria to our simulated QSOs that we obtain the
same colour distribution as for the real data.  This is a necessary,
but not sufficient, requirement to demonstrate that our simulations
accurately model the colours of the underlying population.

Broad-band colours alone, especially when combined with a QSO SED, are
not adequate to fully constrain the host galaxy SEDs.  We therefore
consider a number of possible star-formation scenarios and model SEDs
using the Bruzual \& Charlot (2003) population synthesis code.  We
assume a solar metallicity and Chabrier (2003) initial mass function
(IMF) for all models, on the basis that high redshift QSO
metallicities are typically found to be high (Dietrich et al. 2003;
Kurk et al. 2007), and that with only a small number of broad-band
colours we cannot hope to separate out the effects of any metallicity
or IMF variation.

We first consider two burst models, either a single instantaneous
burst (SIB) or a single long burst of length $\tau=1$Gyr (SLB) which
occurs at high redshift ($z=10$).  We then only allow passive
evolution of the stellar population with redshift.  In such a model,
the normalization (A) in equation \ref{eq:galqsorel} is made relative
to the expected galaxy colours at $z=0$.  

The second set of models we test are those with a fixed age, based on
the argument that QSOs are largely formed in galaxy mergers (e.g. Hopkins
et al. 2006), which will
also trigger star formation.  In this scenario the dominant
stellar population in QSO host galaxies would have similar ages,
consistent with the time since the merger event.   In this scenario we
test three different star formation models, the SIB, the SLB and an
exponential declining (ED) star formation rate with an e-folding
timescale of 1Gyr.  We also test these for a range of ages from 0.5 to
5Gyr.

We ran a suite of simulations for each of these models.  For each
model we fit for the best value 
of $A$ in Eq. \ref{eq:galqsorel} above, while keeping the other
parameters constant.  This accounts for the change in zero-point
due to the differing host SEDs, as well as any host flux missed due to
our use of PSF magnitudes.  We then test the simulated colours
against the 2SLAQ QSO sample to first confirm that they can reproduce
the observed colours.  Once we have ascertained that a given
model is a reasonable description of 2SLAQ colours, we then generate
completeness estimates as a function of redshift and $g-$band magnitude for
all valid simulation parameters.  These results show the
likely range of completeness corrections.

\begin{figure*}
\centering
\centerline{\psfig{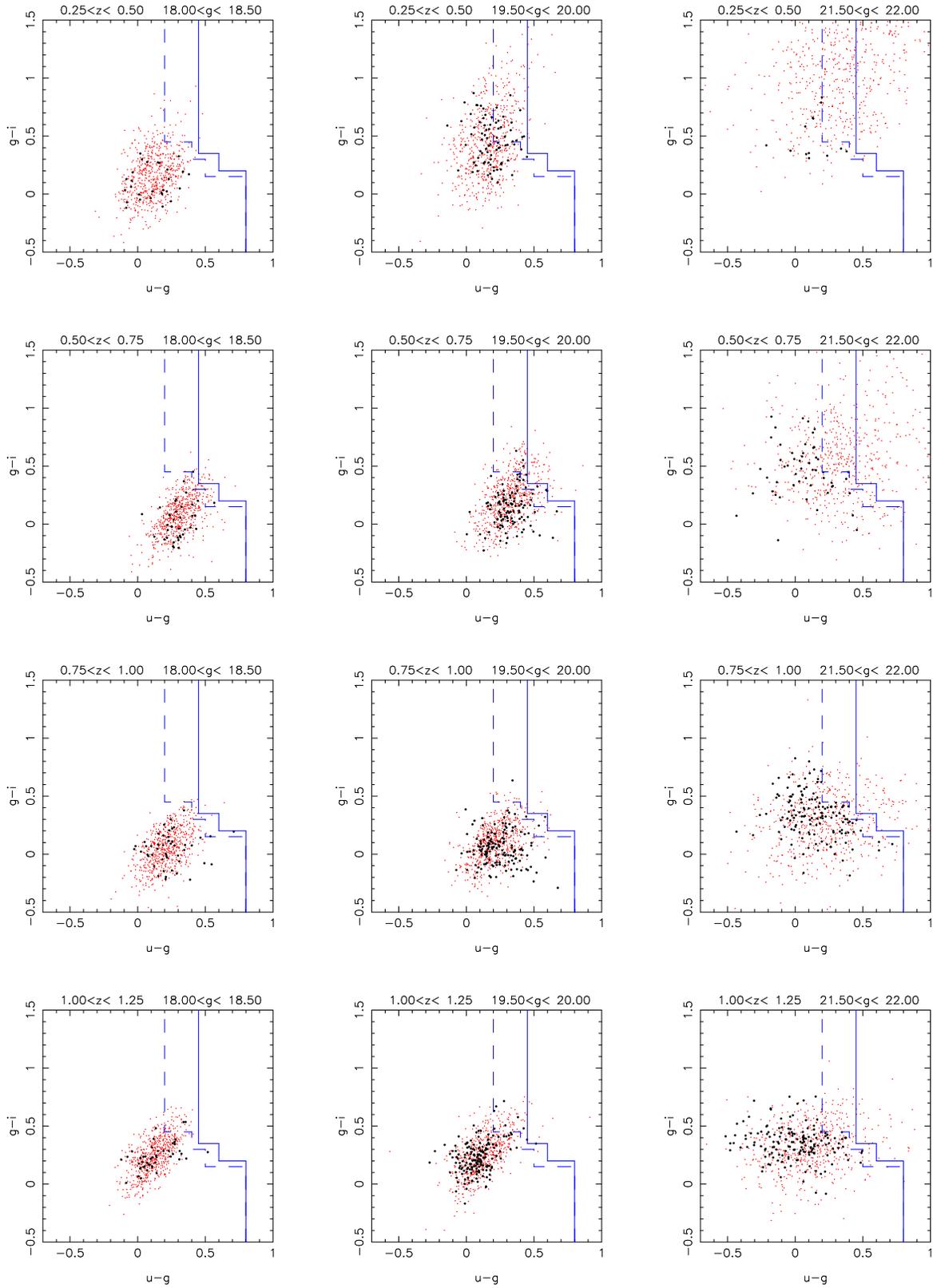}}
\caption{A comparison of 2SLAQ (black points) and simulated (red
  points) $u-g$ and $g-i$ colours in the redshift intervals
  $0.25<z<0.50$,  $0.50<z<0.75$, $0.75<z<1.00$ and $1.00<z<1.25$ (top
  to bottom).  We
  compare these distributions as a function of $g$-band magnitude in
  $\Delta g=0.5$ mag bins with $18.0<g<18.5$, $19.5<g<20.0$ and
  $21.5<g<22.0$ (left to right).  The blue lines
  indicate the QSO selection limits at $g<21.15$ (solid line) and
  $g>21.15$ (dashed line).   Both the data and simulated points show a
  reddening of the $g-i$ colour and increased scatter towards fainter
  magnitudes.  The simulated points are a realization of
  model 2 from Table \ref{tab:colsims}, which provides one of the
  better matches to our data.}
\label{fig:ugi_gz}
\end{figure*}

\begin{table}
\begin{center}
\caption{The parameters and fit results for the different simulation
  parameters used to test 2SLAQ colour completeness.  The simulation
  parameters include the SED model; either single instantaneous burst
  (SIB), single long burst (SLB) or exponential decline (ED).  Also
  listed are the form of evolution and the input slope ($B$) and
  scatter ($\sigma\qg$)
  in Eq. \ref{eq:galqsorel}.  The results of the fit are the best
  value of $A$ and $\chi^2$.  The number of degrees-of-freedom is
  $\nu\simeq68$ in each case.  In two cases, both with fixed ages of
  0.5 Gyr, the fit did not converge, and no best fit value is given.}
\setlength{\tabcolsep}{4pt}
\begin{tabular}{ccccccc}
\hline
\# & SED & Evolution & B & $\sigma\qg$ & A & $\chi^2$\\
\hline
1 & SIB & passive, $\zf=10$ & 0.2 & 0.7 & $-15.53\pm0.03$ & 423  \\
2 & SLB & passive, $\zf=10$ & 0.2 & 0.7 & $-15.51\pm0.03$ & 412  \\
3 & ED  & passive, $\zf=10$ & 0.2 & 0.7 & $-15.44\pm0.03$ & 390  \\

4 & SLB & passive, $\zf=10$ & 0.0 & 0.8 & $-19.99\pm0.03$ & 326  \\
5 & SLB & passive, $\zf=10$ & 0.4 & 0.7 & $-11.05\pm0.03$ & 590  \\

6 & SIB & fixed age 0.5Gyr  & 0.2 & 0.7 & $-16.68\pm0.03$ & 1293 \\
7 & SIB & fixed age 1.0Gyr  & 0.2 & 0.7 & $-16.45\pm0.03$ & 630  \\
8 & SIB & fixed age 3.0Gyr  & 0.2 & 0.7 & $-16.23\pm0.03$ & 485  \\
9 & SIB & fixed age 5.0Gyr  & 0.2 & 0.7 & $-16.20\pm0.03$ & 490  \\

10 & SLB & fixed age 0.5Gyr  & 0.2 & 0.7 & --              & --   \\
11 & SLB & fixed age 1.0Gyr  & 0.2 & 0.7 & $-17.04\pm0.03$ & 2373 \\
12 & SLB & fixed age 3.0Gyr  & 0.2 & 0.7 & $-16.27\pm0.03$ & 470  \\
13 & SLB & fixed age 5.0Gyr  & 0.2 & 0.7 & $-16.21\pm0.03$ & 506  \\

14 & ED  & fixed age 0.5Gyr  & 0.2 & 0.7 &  --             &  --  \\
15 & ED  & fixed age 1.0Gyr  & 0.2 & 0.7 & $-16.54\pm0.04$ & 3027 \\
16 & ED  & fixed age 3.0Gyr  & 0.2 & 0.7 & $-16.65\pm0.03$ & 668  \\
17 & ED  & fixed age 5.0Gyr  & 0.2 & 0.7 & $-16.33\pm0.03$ & 474  \\
\hline
\label{tab:colsims}
\end{tabular}
\end{center}
\end{table}

\begin{figure*}
\centering
\centerline{\psfig{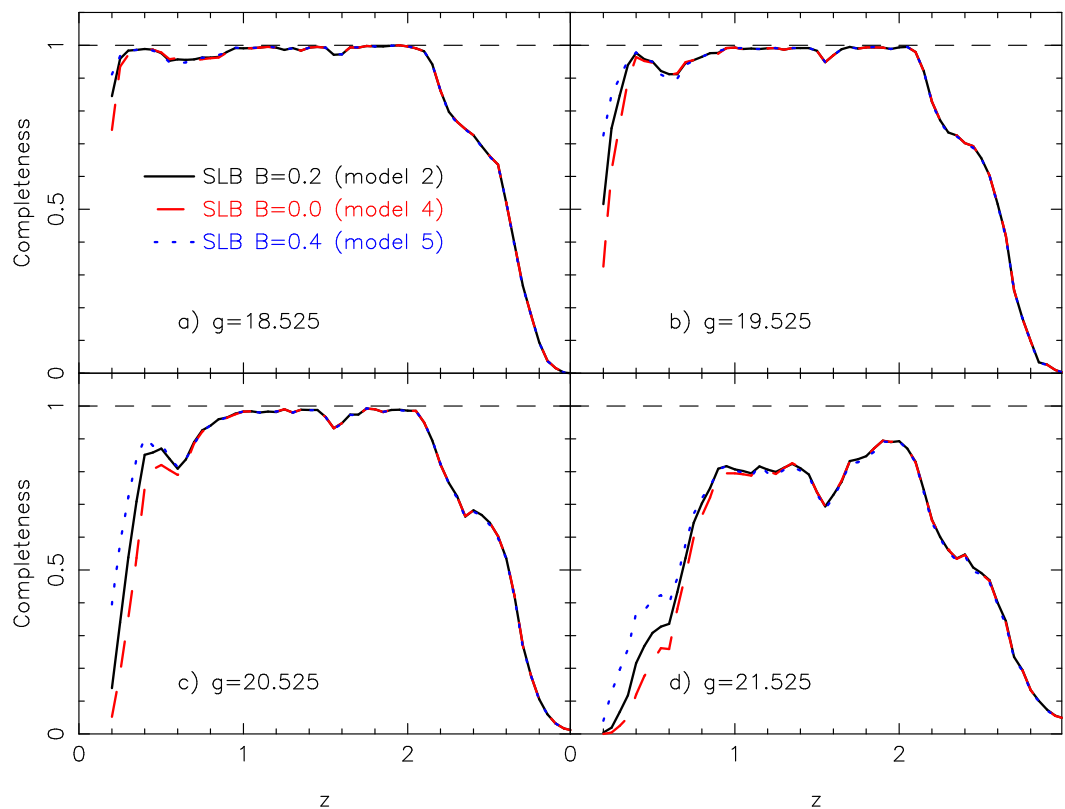}}
\caption{A comparison of photometric completeness as a function of
  redshift for 4 different apparent magnitude slices with mean
  $g$-band mag a) $g=18.525$, b) $g=19.525$, c) $g=20.525$ and d)
  $g=21.525$.  We compare models with different values of $B$ (defined
  in Eq. \ref{eq:galqsorel}), $B=0.0$
  (dashed red line), $B=0.2$ (solid black line), $B=0.4$ (dotted blue
  line) which are models 4, 2 and 5 respectively in Table
  \ref{tab:colsims}.  At bright magnitudes and/or high redshift all
  three models agree very well, while at faint magnitudes the models
  diverge below $z\sim0.5$.}
\label{fig:colcomp_slice}
\end{figure*}

Table \ref{tab:colsims} contains the parameters for the various host
galaxy models used in the above tests.  An example of the colour
distributions as a function of $g$ and redshift is shown in
Fig. \ref{fig:ugi_gz}, which compares simulated colours for model 2
and observed 2SLAQ colours.  To test each model we 
sample the redshift distribution from $z=0.1$ to 1.5 and total magnitude
(nucleus + host) distribution from $g=18$ to 22, binning with $\Delta
z = 0.02$ and $\Delta g = 0.1$.  In each of these bins we generate 10
model QSO+host spectra, and then compare the derived $g-i$ colours to
those from the 2SLAQ sample in order to obtain the best fit value of
$A$ from Eq. \ref{eq:galqsorel} above.  The best fit values and the
resulting $\chi^2$ are listed in table \ref{tab:colsims}.  The $g-i$
colour comparison was made with the median colours in bins of $\Delta
z=0.25$ and $\Delta g=0.1$.  The errors on these median values were
taken as the 68 percent inter-quartile value.  Only bins with 5 or
more observed and simulated points were considered.  The number of
degrees of freedom was $\nu\simeq68$ in each case, varying slightly
due to our constraint of requiring at least 5 points in each bin.  It
can be seen that all the $\chi^2$ values are considerably larger than
$\nu$, indicating that in detail our relatively simple model does not
perfectly trace the distribution of QSO+host colours.  We note here
that we are not aiming to model the host colours in great detail, but
to obtain a description of them that is sufficiently accurate to enable a
reasonable estimate of completeness.

For a model with passive evolution and an early
redshift of formation ($\zf=10$) there is little difference between
the models; largely because at the redshift in question, all the
models SEDs are dominated by older stars.  When we change the slope of
the $M\gal-M\qso$ relation (Eq. \ref{eq:galqsorel}), a
flatter slope is preferred according to our $\chi^2$ statistic.  This
is because a flatter slope in Eq. \ref{eq:galqsorel} allows the
relation between $g-i$ and $g$ to be steeper.  However, we note that
if we push the slope to be even flatter (i.e. $B\leq0$), then although we
obtain a relatively good fit with the test described above, the simulated
objects at low $z$ and faint $g$ are too red.  This discrepancy does not
show up in the above test because of the relatively low number of
2SLAQ objects in these bins.

Considering the models with fixed age SEDs (numbers 6--17 in Table
\ref{tab:colsims}), it is clear that models 
with young SEDs are very much worse fits than older SEDs.  The
$\chi^2$ values increase substantially for the SIB and SLB models with
ages $<3$Gyr.  For the ED model, only the 5Gyr population gives as
good a fit as the SIB and SLB models.  The data are
more consistent with an older age stellar population in the
host galaxies because of the red colours of faint 2SLAQ
objects in $g-i$.  This can be seen in Fig \ref{fig:ugi_gz}, which
shows the reddening of $\sim0.5$ mag in $g-i$ from $g=18.0$ to 22.0,
in the redshift interval $0.50<z<0.75$.
A significant young and blue ($<1$Gyr) stellar population is
inconsistent with the observed trend, as reflected in in the poorer
$\chi^2$ fits in Table \ref{tab:colsims}.  This could provide challenges for models
of QSO formation which rely on mergers to trigger QSO activity and
star formation \cite{hop06a}, as it implies that the dominant stellar
population is relatively old.  However, given that the objects in
question are at low redshift and low luminosity, it could be that
major mergers do not play such a significant role for this faint
population \cite{hophern06}.  Indeed, various observations of more
luminous QSO have found evidence that the hosts are bluer than typical
galaxies, with relatively young stellar populations (e.g. Kirhakos et
al. 1999; Zakamska et al. 2006).
At low redshift ($z<0.3$) Kauffman et al. (2003) show that the stellar
populations in the host galaxies of type II AGN are similar to
non-active early type galaxies, although at high luminosity the
stellar population becomes younger (as measured by the $D_{\rm
  n}(4000)$ index).  Vanden Berk et al. (2006) find that hosts of SDSS
quasars at $z<0.75$ span a wide range of spectral types, but at low
host (and nuclear) luminosity the typical colours are similar to
non-active early types.  Our estimates of host galaxies properties
seem broadly consistent with this work.

\begin{figure}
\centering
\centerline{\psfig{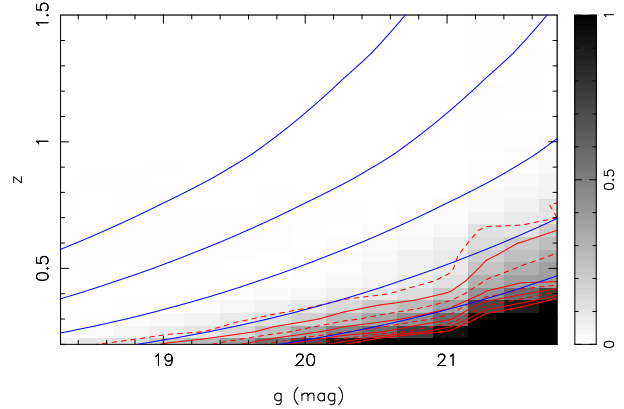}}
\caption{The fractional difference between completeness estimates
  assuming host galaxy models 2, 4 and 5.  We plot $0.5(C_4-C_5)/C_2$
  where $C_4$ is the completeness in model 4 etc.  The contours are at 0.1
  intervals (0.1, 0.3... dashed red lines; 0.2, 0.4... solid red
  lines), and the same data are also plotted as a grey-scale.  The
  solid blue lines denote lines of constant absolute magnitude from
  $\mg=-24$ (top) to $\mg=-20$ (bottom) in 1.0 mag intervals.  At
  $\mg=-21$ the uncertainty in the completeness is always less than 20
  per cent.  Note the more limited redshift range compared to
  Fig. \ref{fig:colcomp} below.}
\label{fig:colcomperr}
\end{figure}

The final step is to compare the completeness estimates from the
different models in order to determine the uncertainty in our
completeness corrections.  We simulate the redshift range $0.1<z<3.0$ and use $18<g<22$.
We use the bins $\Delta z = 0.02$ and $\Delta g = 0.1$ and increase
the number of simulated objects to 200 per bin, so that 
our completeness estimates are not limited by shot-noise.  We find
that the largest variation in estimated completeness from within our
suite of models is due to the variation in the parameter $B$ in Eq.
\ref{eq:galqsorel}.  In Fig. \ref{fig:colcomp_slice} we plot models 2,
4 and 5 from Table \ref{tab:colsims} for four different $g$-band magnitude
slices.  It is only at faint magnitudes and redshifts less than
$\sim0.5$ that the completeness depends on the model.  To quantify
this further, Fig. \ref{fig:colcomperr} plots the mean fractional difference in the models,
i.e. $0.5*(C_4-C_5)/C_2$, where $C_2$ is the completeness in model 2
etc.  We overlay contours
of constant absolute $g$-band magnitude (blue lines) from $\mg=-24$
(top) to $-20$ (bottom) (using the Cristiani \& Vio 1990
K-correction).  Brighter than 
$\mg\simeq-21$ the uncertainty on the completeness estimate is less
than 20 per cent and by $\mg\simeq-21.5$ this uncertainty is less than 5
per cent.

\begin{figure}
\centering
\centerline{\psfig{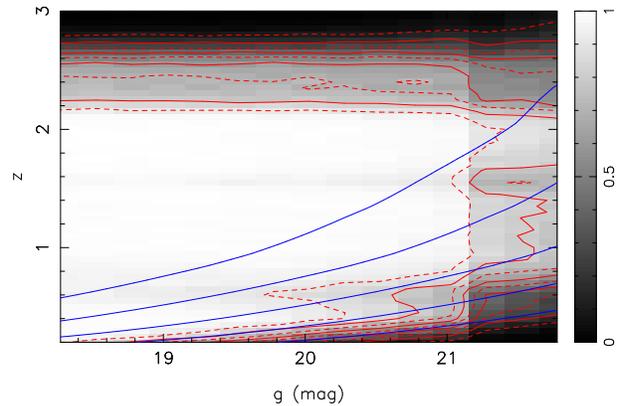}}
\caption{The final photometric completeness map for the 2SLAQ sample as
  a function of $g$ and $z$ assuming model 2.  The contours are at
  0.1 intervals (0.1, 0.3... dashed red lines; 0.2, 0.4... solid red
  lines), and the same data are also plotted as a grey-scale.  The
  solid blue lines denote lines of constant absolute magnitude from
  $\mg=-24$ (top) to $\mg=-20$ (bottom) in 1.0 mag intervals.}
\label{fig:colcomp}
\end{figure}

\begin{figure}
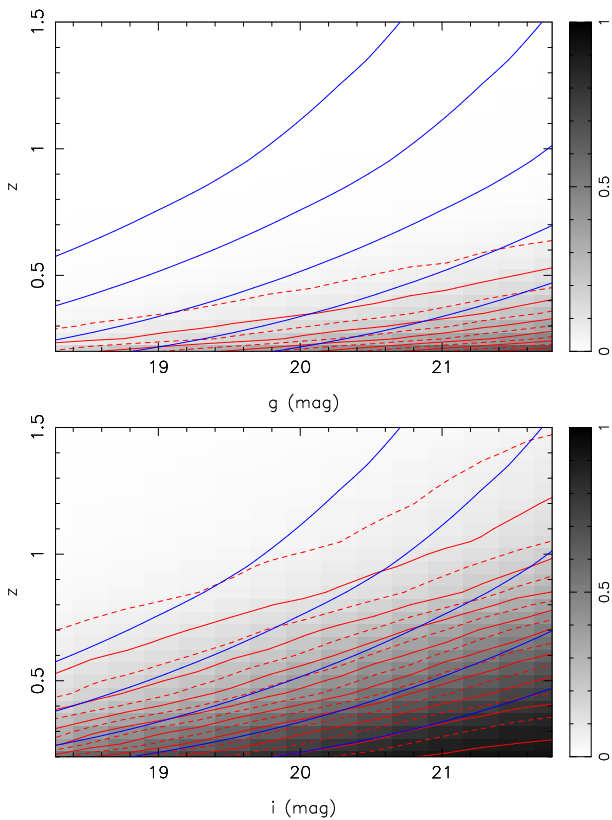

\centering
\centerline{\psfig{file=host_frac_g.ps,width=8cm,angle=270}}
\centerline{\psfig{file=host_frac_i.ps,width=8cm,angle=270}}
\caption{Top: the fractional contribution of the host galaxy to the
  total $g$-band PSF flux of 2SLAQ QSOs as a function of $g$ and $z$
  (assuming model 2 from Table \ref{tab:colsims}).
  The contours are at 0.05 intervals (0.05, 0.15... dashed red lines;
  0.1, 0.2... solid red lines), and the data are also plotted as a 
  grey-scale.  The solid blue lines denote lines of constant absolute
  magnitude from $\mg=-24$ (top) to $\mg=-20$ (bottom) in 1.0 mag
  intervals. Bottom: same as above, but showing the fractional
  contribution of the host galaxy to the total $i$-band PSF flux as a
  function of redshift and $i$-band flux.  At low
  redshift and faint $i$-band flux, these $i$-band host contributions
  are extrapolated from the comparison to the 2SLAQ data.  This is
  because the combined QSO+host colours are relatively red,
  $g-i\sim1$, so that a source with $i\simeq22$ will be
  $\sim1$ mag fainter than the 2SLAQ $g$-band limit.}
\label{fig:hostfrac} 
\end{figure}

\begin{table}
\begin{center}
\caption{The 2SLAQ colour completeness array.  The completeness is
  given as a function of redshift, $z$, and $g$-band magnitude.  We
  also list the host galaxy contribution as a magnitude difference,
  $\Delta$host in both the $g$ and $i$ bands.  The first ten rows are
  given here.  The full table is available in the electronic version
  of the journal.}
\setlength{\tabcolsep}{4pt}
\begin{tabular}{ccccc}
\hline
 $z$ & Mag. & Completeness & $\Delta$host & $\Delta$host\\
     & $g$/$i$ &           &   $g$ (mag)  & $i$ (mag)\\
\hline
  0.200 & 18.275 & 0.92 & 0.19 & 0.81\\
  0.200 & 18.525 & 0.85 & 0.23 & 0.99\\
  0.200 & 18.775 & 0.79 & 0.27 & 1.11\\
  0.200 & 19.025 & 0.69 & 0.31 & 1.26\\
  0.200 & 19.275 & 0.62 & 0.36 & 1.41\\
  0.200 & 19.525 & 0.52 & 0.40 & 1.55\\
  0.200 & 19.775 & 0.40 & 0.46 & 1.73\\
  0.200 & 20.025 & 0.31 & 0.53 & 1.91\\
  0.200 & 20.275 & 0.23 & 0.60 & 2.10\\
  0.200 & 20.525 & 0.14 & 0.68 & 2.29\\
\hline
\label{tab:colcomp}
\end{tabular}
\end{center}
\end{table}

Based on this we adopt model 2 for our final colour completeness.
This provides a reasonable match to the observed 2SLAQ QSO colours as
well as being mid-way between the models with the 
largest variations (i.e. models 4 and 5).  The completeness array for
this model is shown in Fig. \ref{fig:colcomp}.  The step at $g=21.15$,
where we change our colour selection limits, is clearly visible.  In
contrast to our previous estimates of the completeness of 2SLAQ
selection (see R05), the completeness derived here declines much more
at faint magnitudes and low redshift. 
Table \ref{tab:colcomp} contains the colour completeness array as a
function of $z$ and $g$.  The full table is available in the
electronic version of the journal.

\subsubsection{Correcting for host galaxy flux}

Our detailed simulations enable a relatively straightforward mechanism
to correct the photometry of the 2SLAQ QSOs for
the contribution of their host galaxy.  The photometry used to select
targets and calculate luminosities was based on SDSS PSF magnitudes,
so this should limit the contribution of the host to some extent, but
up to $z\simeq1$ there is some contribution from the host in the PSF
magnitudes.  For each simulated source we calculate the total PSF
magnitude ($m_{\rm tot}$, QSO+host) and the difference 
between total and the QSO, $m_{\rm QSO}-m_{\rm tot}$.  We can then
derive the mean correction from total to nuclear magnitude in the same
$g$-band and redshift intervals used to make the completeness correction
array.  Fig. \ref{fig:hostfrac} shows the fractional host
contribution to the PSF flux in the $g$ (top) and $i$ (bottom) bands.
In the $g$-band, which forms the flux limit of the 2SLAQ sample, the
host contribution is less than 20 per cent at $z>0.4$, even
for the faintest sources.  However, it is not until
$z\sim0.9$ that the contribution falls below 20 per cent in the
$i$-band.  These corrections can be applied to determine nuclear
fluxes for 2SLAQ sources.  The host galaxy contributions as a
function of redshift and $g$ and $i$ band magnitude are listed in Table
\ref{tab:colcomp}.

\subsection{Coverage completeness}

The coverage completeness in the 2SLAQ survey is a function of both
celestial position and $g$-band magnitude.  The global coverage as 
a function of $g$ is shown in Fig. \ref{fig:covcomp}.  The step at
$g=20.5$ is due to the prioritization of faint QSOs, while the
increasing completeness from $g=20.5$ towards bright magnitudes is due
to our inclusion of observations from the 2QZ and SDSS surveys.  This
global coverage completeness is all that is required for 
analyses such as luminosity function calculations.

The angular dependence on the sky of the coverage is a fixed value for
each sector made up by the intersection of overlapping 2SLAQ fields.
However, the geometry of these fields is more complex than in the case
of 2QZ, as we must account for the triangular exclusion regions
around the edge of each field (Fig. \ref{fig:fldcomp}).  Once these
areas are accounted for we are able to construct a completeness mask
with $1\times1$ arcmin pixels which describes the angular coverage of
the survey.

\begin{figure}
\centering
\centerline{\psfig{file=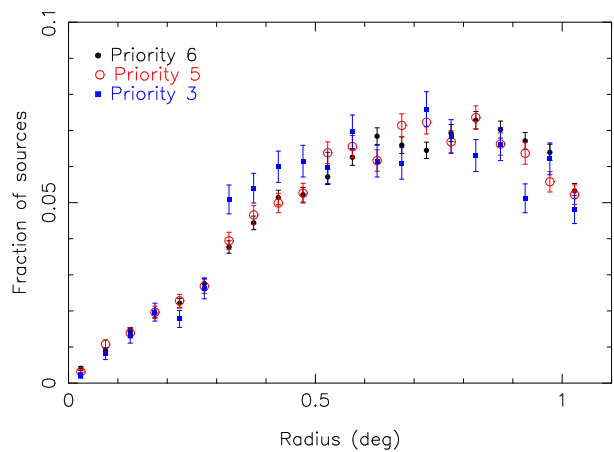,width=8cm,angle=270}}
\caption{The fraction of 2SLAQ objects targetted spectroscopically as a function of
  radius from the 2dF field centre in bins of $\Delta r=0.05^\circ$.
  We plot the radial distribution for the priority 6 (black filled
  circles), 5 (red open circles) and 3 (blue filled squares).}
\label{fig:radialdiff}
\end{figure}

A second complication is that because of the biases in the {\small
CONFIGURE} software [see Section \ref{sec:targpri} above and Mizarksi
et al. (2006)] it is possible that targets at different priorities
could be distributed differently within a 2dF field.  We test this by
comparing the spatial distribution of our highest priority QSOs
(priority 6; see Table \ref{tab:pri}) and the priority 5 and 3 QSOs
(there are too few priority 4 QSOs to make a meaningful comparison).  We carry out
two tests.  The first is to bin the observed objects as a function of
radius from their field centre, with $\Delta r=0.05^\circ$ (see
Fig. \ref{fig:radialdiff}).  A $\chi^2$ test between priority 6 and 5
objects gives $\chi^2/\nu=24.17/21$ which is only inconsistent at the
28 per cent level.  By contrast comparing priority 6 and 3 we find
$\chi^2/\nu=54.80/21$, which infers a probability of being drawn from the
same population of $7.56\times10^{-5}$.
We also compare the distributions using a 2D KS test, where the two
dimensions correspond to the angular coordinates ($\alpha$, $\delta$)
of the QSOs on the sky.  For the 
priority 6/5 comparison this gives $D_{\rm KS}=0.0224$ and $P(>D_{\rm
KS})=0.11$, while for the priority 6/3 comparison we find $D_{\rm
KS}=0.103$ and $P(>D_{\rm KS})=1.013\times10^{-15}$.  Thus we conclude
that there is no statistically significant difference in the spatial distribution of priority 5
and 6 objects, but that the priority 3 sources (bright, $g<20.5$ QSOs)
do have a significantly different spatial distribution.  These objects
should then not be used for analyses in which spatial distribution is
important (e.g. clustering measurements).

\begin{figure}
\centering
\centerline{\psfig{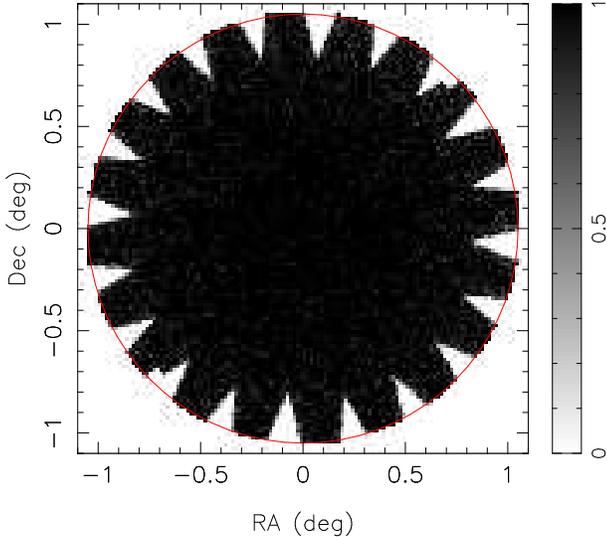}}
\caption{The 2SLAQ QSO coverage mask for a single 2dF field,
  determined by averaging over many configurations of random
  simulations.  The greyscale gives the probability of being able to
  configure an object.  The mask is defined in $1\times1$ arcmin pixels, and
  the plotted (red) circle defines the $1.05^\circ$ radius field-of-view.
  The inaccessible wedges around the edge of the field are clearly
  visible.  The small holes near the bottom left ($\alpha=\delta=-0.7^\circ$)
  and top right ($\alpha=\delta=+0.7^\circ$) of the
  field are due to two retaining screws in the field plate, over which
  fibres cannot be placed.}
\label{fig:fieldmask}
\end{figure}

\begin{figure*}
\centering
\centerline{\psfig{file=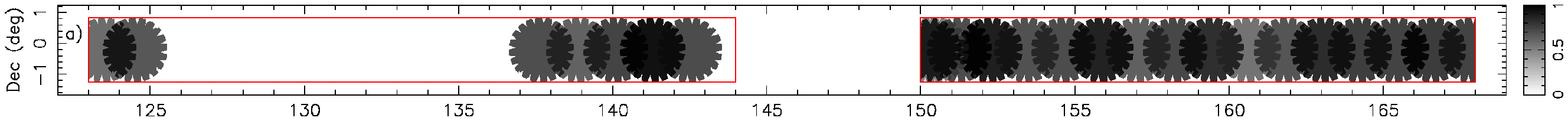,width=18cm,angle=270}}
\vspace{0.1cm}
\centerline{\psfig{file=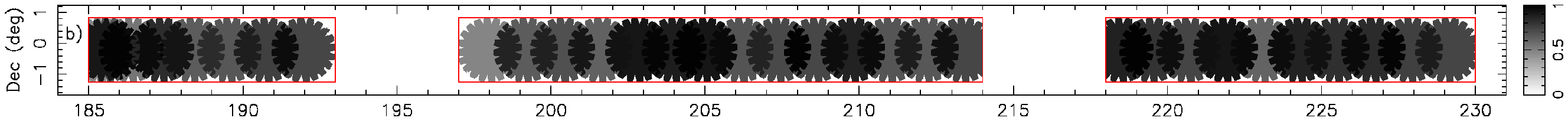,width=18cm,angle=270}}
\vspace{0.1cm}
\centerline{\psfig{file=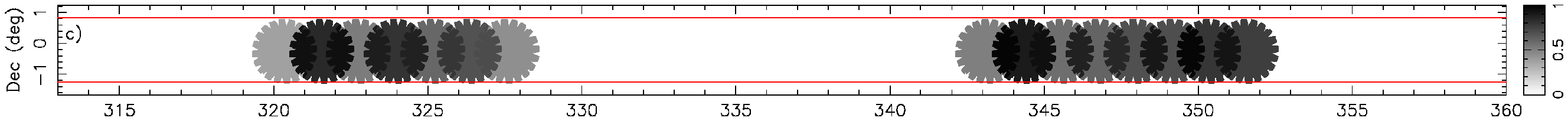,width=18cm,angle=270}}
\vspace{0.1cm}
\centerline{\psfig{file=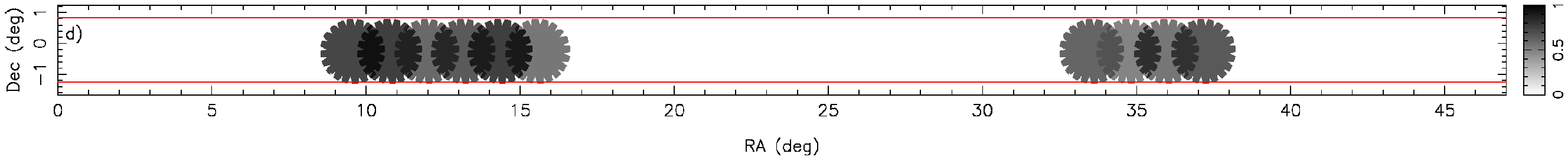,width=18cm,angle=270}}
\caption{The 2SLAQ coverage mask, defined as the fraction of priority
  5 and 6 targets (QSO candidates with $g>20.5$) that  have been
  spectroscopically observed.  The grey-scale extends from 0.0 (white)
  to 1.0 (black).  The solid red lines mark the extent of the input
  catalogue regions. a) The a and b regions in the NGP; b) the c, d
  and e regions in the NGP, c/d) the single s region in the SGP.
  While there is almost complete coverage in the NGP regions (apart
  from the a region at $\alpha = 123$ to $144$ deg), large parts of
  the SGP region remain unobserved.}
\label{fig:covmask}
\end{figure*}

\begin{figure*}
\centering
\centerline{\psfig{file=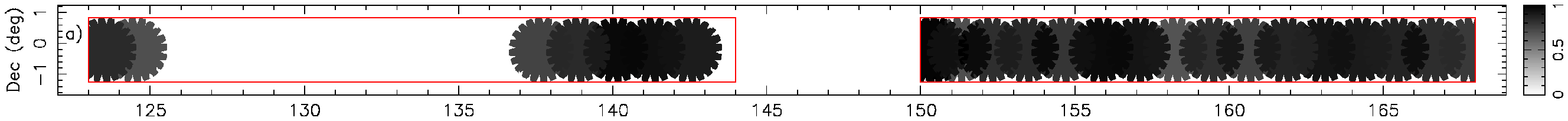,width=18cm,angle=270}}
\vspace{0.1cm}
\centerline{\psfig{file=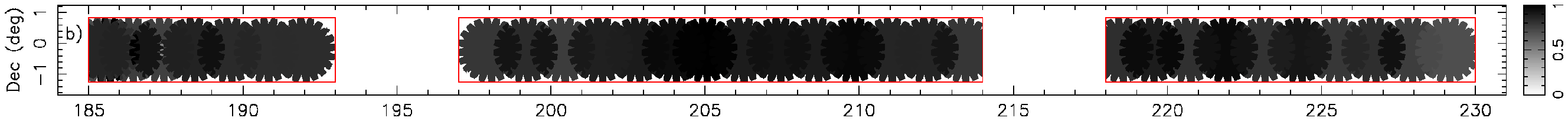,width=18cm,angle=270}}
\vspace{0.1cm}
\centerline{\psfig{file=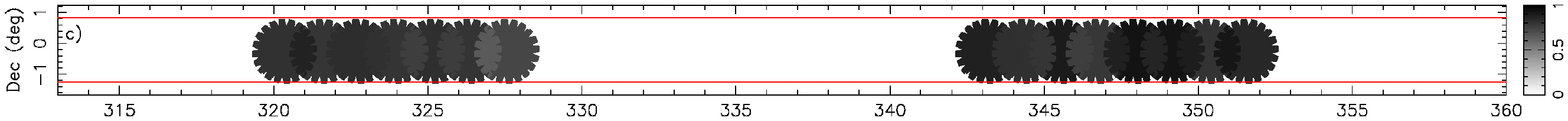,width=18cm,angle=270}}
\vspace{0.1cm}
\centerline{\psfig{file=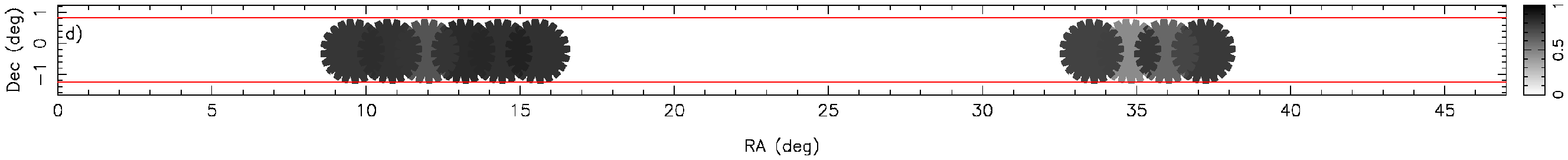,width=18cm,angle=270}}
\caption{The 2SLAQ spectroscopic completeness mask, defined as the
  fraction of observed priority 5 and 6 objects which have high
  quality identifications (ID quality = 1).  The grey-scale extends
  from 0.0 (white) to 1.0 (black).  The regions plotted are the same
  as in Fig. \ref{fig:covmask}.}  
\label{fig:specmask}
\end{figure*}

We derive the average mask for a single 2dF pointing, taking
into account the inaccessible wedges near the edge of the field.  This
is done by running the {\small CONFIGURE} software in batch mode for
1000 realizations.   The resulting single field
mask is shown in Fig. \ref{fig:fieldmask}.  This is then converted to
a binary mask (i.e. 1 if observable, or 0 if not) to define the
boundaries for each field.  With the geometry of each field defined,
we determine a set of unique sectors formed from the overlap of all
the observed 2SLAQ fields.  Within each sector we then derive the
fraction of priority 5 and 6 objects observed.  This is then sampled
onto $1\times1$ arcmin pixels.  The resulting coverage masks for the
NGP and SGP regions are shown in Fig. \ref{fig:covmask}.  As this mask
is for the priority 5 and 6 objects only, it has no 
magnitude dependence and it is a function of
angular celestial position only, i.e. $\fc(\alpha,\delta)$.  Another
issue to consider when studying the spatial distribution of the 2SLAQ
QSOs is that the 2SLAQ LRGs were given higher priority.  Thus the
regions within $\simeq30$ arcsec of the LRGs (Cannon et al. 2006) represent regions of sky
that were not surveyed, and these need to be included in the coverage
mask. This has been done for the QSO clustering analysis presented by da
Angela et al. (2008).  The masks presented in this current work are
not corrected for the LRG distribution.

\subsection{Spectroscopic completeness}

We specify the spectroscopic completeness as the ratio 
$N_1/N_{\rm obs}$, where $N_1$ is the number of quality 1 IDs and
$N_{\rm obs}$ is the number of targets observed.  The global
spectroscopic completeness is shown in Fig. \ref{fig:objfrac}b as a
function of $g$-band mag only.  In general
this is a function of angular position (due to varying observing
conditions etc.) and redshift (due to different emission lines moving
in and out of the observed spectral range) as well as $g$.  We
make the assumption that the fraction of QSOs among the unidentified
objects is the same as that within the sample of high quality
identifications.  We saw that this is reasonable from our analysis of
repeat observations in Section
\ref{sec:rep}.  Below we follow C04 to generalize the spectroscopic
completeness estimate to be a function of celestial position
($\alpha,\delta$), $g$-band magnitude and redshift.

\subsubsection{Position-dependent spectroscopic completeness}

We can generate a position dependent spectroscopic completeness mask,
$\fs(\alpha,\delta)$, similar to the coverage mask presented above.  The mean
spectroscopic completeness per sector is shown in
Fig. \ref{fig:specmask}.  Because the observations were
extended during poor observing conditions, the spectroscopic
completeness is relatively uniform.

\subsubsection{Magnitude-dependent spectroscopic completeness}

\begin{figure}
\centerline{\psfig{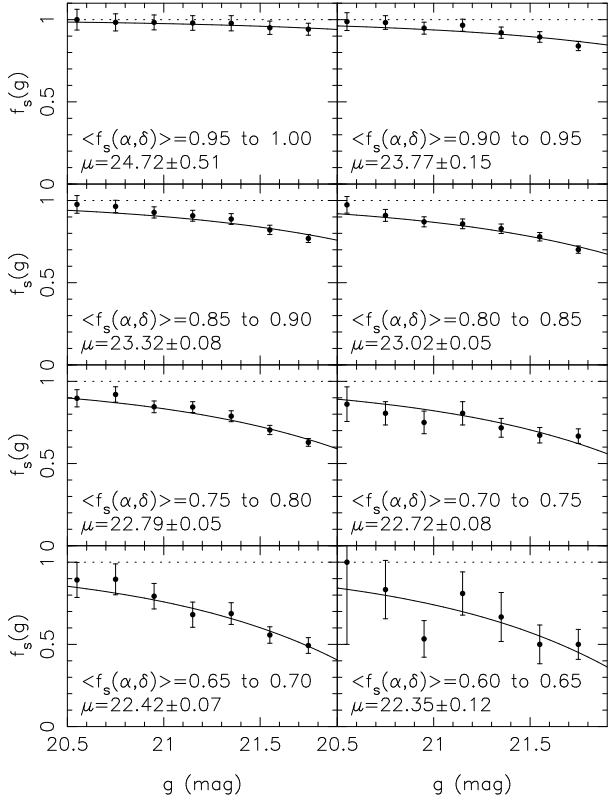}}
\caption{The 2SLAQ spectroscopic completeness as a function of
  $g$-band magnitude for sectors with different mean completeness
  values, $\langle\fs(\alpha,\delta)\rangle$.  In
  each case the best fit magnitude-dependent completeness model
  (Eq. \ref{eq:mumodel}) is shown (solid line).}
\label{fig:fsmag}
\end{figure}

\begin{figure}
\centerline{\psfig{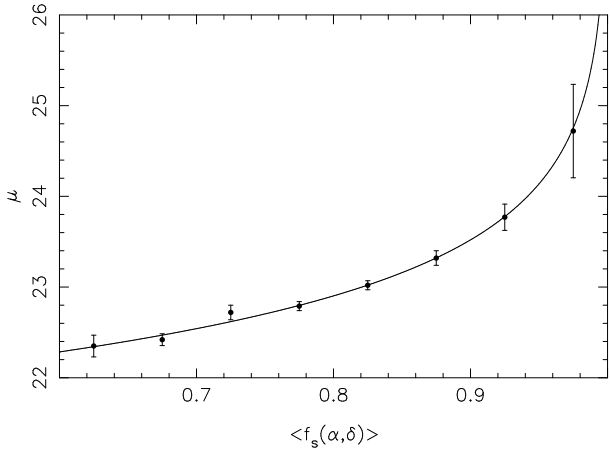}}
\caption{The magnitude dependent spectroscopic completeness parameter,
  $\mu$, as a function of mean sector spectroscopic completeness,
  $\langle\fs(\alpha,\delta)\rangle$ (filled circles).  The solid line shows the best fit
  of Eq. \ref{eq:mudep} to this data.}
\label{fig:mu}
\end{figure}

Some analyses, e.g. luminosity dependent clustering measurements, will
require that the magnitude dependence of the completeness variations are
accurately mapped over the 2SLAQ regions.  Following C04, we determine
the magnitude dependent spectroscopic completeness, $\fs(g)$, within
sectors of varying completeness.  This is plotted in
Fig. \ref{fig:fsmag}.  For fields with high
average completeness, there is very little magnitude dependence, but
the magnitude dependence becomes stronger as
the average completeness declines.  We parameterize the magnitude
dependence of completeness by the function
\begin{equation}
\fs(g)=1-\exp(g-\mu),
\label{eq:mumodel} 
\end{equation}
where $\mu$ is a parameter fit in each magnitude bin.  The best fits of this
function for each average completeness interval are plotted in
Fig. \ref{fig:fsmag}.  The $\mu$ parameter varies smoothly with
average completeness (see Fig. \ref{fig:mu}) and can be well described
by 
\begin{equation}
\mu=A+B\ln[1-\langle\fs(\alpha,\delta)\rangle],
\label{eq:mudep}
\end{equation}
where $A=21.47\pm0.11$ and $B=-0.891\pm0.070$.  The magnitude dependence of
$\fs$ in a particular sector is then described by
\begin{equation}
\fs(g)=1-\frac{\exp(g-A)}{[1-\fs(\alpha,\delta)]^B}.
\end{equation}
This can be combined  with $\fs(\alpha,\delta)$ to give
\begin{equation}
\fs(\alpha,\delta,g)=\frac{N_{\rm obs}(\alpha,\delta)}{N_{\rm est}(\alpha,\delta)}\fs(\alpha,\delta)\fs(g),
\end{equation}
where the value $N_{\rm est}(\alpha,\delta)$ is the estimated number of
quality 1 IDs given the function $\fs(g)$ in a particular sector.  This
is given by
\begin{equation}
N_{\rm est}(\alpha,\delta)=\sum^{N_{\rm obs}(\alpha,\delta)}_{i=1} \fs(g).
\end{equation}
The angular masks containing the values of $\fs(\alpha,\delta)$ and $N_{\rm
  obs}(\alpha,\delta)/N_{\rm est}(\alpha,\delta)$ are supplied as part of the 2SLAQ
  catalogue release.

For analyses which do not require spatial information (e.g. the
luminosity function of QSOs), a global average best fit $\mu$ value can
be used.  For the priority 5 and 6 sample discussed here we find a
global best fit $\mu=23.21\pm0.03$.  The average completeness of the
sample is 85.4 per cent, which would imply $\mu=23.18$ using the best
fit values of $A$ and $B$ above, in good agreement with the directly
fitted value.

\subsubsection{Redshift-dependent spectroscopic completeness}  

\begin{figure}
\centering
\centerline{\psfig{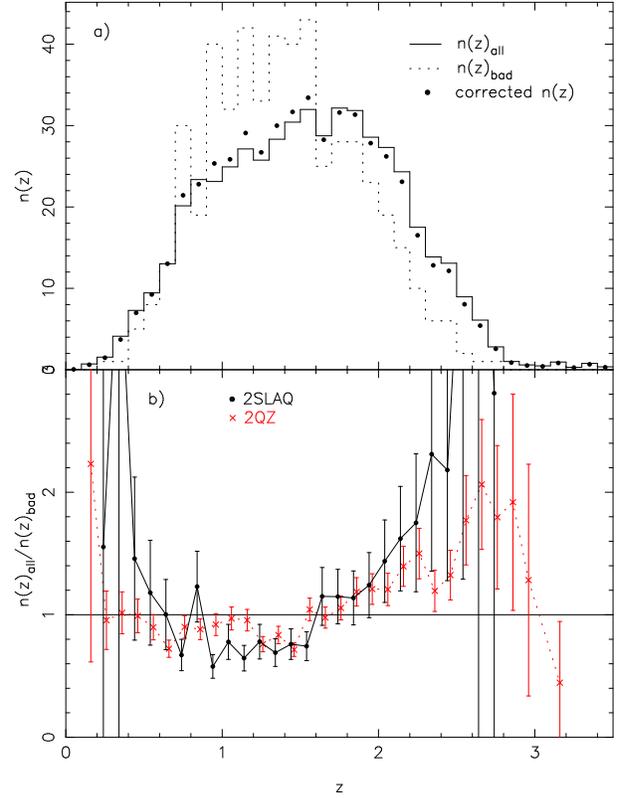}}
\caption{The redshift dependence of spectroscopic completeness in the
  2SLAQ survey. a) The observed distribution of all quality 11 2SLAQ
  QSOs, $n(z)\all$ (solid line), compared to the redshift distribution
  of objects with low-quality identifications which also obtained
  high-quality QSO redshifts, $n(z)\bad$ (dotted line).  $n(z)\all$
  has been renormalized to match $n(z)\bad$.  Also plotted is the
  corrected $n(z)$ assuming that unidentified objects are distributed
  as $n(z)\bad$. b)  The ratio of the normalized $n(z)\all$ to
  $n(z)\bad$ with Poisson error bars (connected filled circles)  As a
  comparison, the same quantity derived from the 2QZ sample is also
  shown (red crosses).}
\label{fig:zspeccomp}
\end{figure}

We follow a procedure similar to that employed by the 2QZ
to determine if there are any spectroscopic completeness variations as
a function of redshift.  To do this we use the 512 QSOs with good
quality (11) IDs, that also have repeat observations of lower quality
(22 or worse).  We can use these repeated observations to examine the
redshift distribution of unidentified objects.  In
Fig. \ref{fig:zspeccomp}a we plot the redshift distribution of all
quality 11 QSOs observed as part of 2SLAQ (solid line), $n(z)\all$,
compared to the redshift distribution of repeated objects that had at
least one bad ID as well as a good ID (dotted line), $n(z)\bad$.  The
fraction of bad IDs is higher in the redshift range $z\simeq0.5-1.6$,
probably due to the changing visibility of strong emission lines
as a function of redshift.  In particular, the \civ\ emission line
becomes visible in the spectra at $z\sim1.5$ making identifications
significantly easier above this redshift.  The ratio of $n(z)\all$ to
$n(z)\bad$ is shown in Fig. \ref{fig:zspeccomp}b (filled circles).
If we then add in the 13.1 per cent of objects without good
quality IDs in the 2SLAQ catalogue (assuming they have the same
redshift distribution as $n(z)\bad$), we get the redshift distribution
given by the filled points in Fig. \ref{fig:zspeccomp}a.  Thus, even
though the redshift distribution of unidentified objects is 
quite different from the rest of the sample, the small
incompleteness means that the overall $n(z)$ is changed only
slightly.  As a comparison we also plot the ratio of $n(z)\all$ to
$n(z)\bad$ derived from the 2QZ (C04) (red crosses in
\ref{fig:zspeccomp}b).  The 
2QZ and 2SLAQ samples show exactly the same trends, although the
amplitude of the variations is somewhat higher for the 2SLAQ sample.
This could be due to the Baldwin (1977) effect, which causes emission lines to
be stronger in lower luminosity QSOs.

\begin{table}
\centering
\caption{Derived values for $R(z)\equiv n(z)\all/n(z)\bad$ (and errors, $\sigma_{R(z)}$)
that can be used to estimate the redshift dependence of spectroscopic
incompleteness.  Data are binned in $\Delta z=0.1$ bins.}
\label{tab:fsz}
\begin{tabular}{rrrrrr}
\hline
$z$ & $R(z)$ & $\sigma_{R(z)}$ & $z$ & $R(z)$ & $\sigma_{R(z)}$\\ 
\hline 
 0.15 &  0.72 &  0.75 & 1.55 &  0.74 &  0.12\\
 0.25 &  0.78 &  0.57 & 1.65 &  1.15 &  0.24\\
 0.35 &  4.13 &  4.16 & 1.75 &  1.15 &  0.22\\
 0.45 &  1.46 &  0.67 & 1.85 &  1.14 &  0.22\\
 0.55 &  1.05 &  0.36 & 1.95 &  1.24 &  0.27\\
 0.65 &  1.00 &  0.29 & 2.05 &  1.44 &  0.34\\
 0.75 &  0.67 &  0.13 & 2.15 &  1.62 &  0.42\\
 0.85 &  1.17 &  0.27 & 2.25 &  1.75 &  0.56\\
 0.95 &  0.58 &  0.10 & 2.35 &  2.31 &  0.96\\
 1.05 &  0.78 &  0.14 & 2.45 &  2.18 &  0.90\\
 1.15 &  0.65 &  0.10 & 2.55 &  4.48 &  3.19\\
 1.25 &  0.78 &  0.14 & 2.65 &  6.10 &  6.13\\
 1.35 &  0.69 &  0.11 & 2.75 &  2.81 &  2.84\\
 1.45 &  0.76 &  0.13\\
\hline
\end{tabular}
\end{table} 

Following the procedure used for the 2QZ, we define a function
$R(z)$ such that
\begin{equation}
R(z)=\frac{n(z)\all}{n(z)\bad},
\end{equation}
where both $n(z)\all$ and $n(z)\bad$ are normalized such
that the sum of all redshift bins is equal to one.  This is the value
plotted in Fig. \ref{fig:zspeccomp}b and also tabulated in Table
\ref{tab:fsz}.  The angular and magnitude dependent 
completeness is then given by
\begin{equation}
\fs(\alpha,\delta,g,z)=\left\{1+\frac{1}{R(z)}\left[\frac{1}{\fs(\alpha,\delta,g)}-1\right]\right\}^{-1}.
\end{equation}  
This relation can then be used to describe the variations in
completeness in a general sense for the 2SLAQ sample.  We note that in
this analysis we have assumed that the redshift
dependence is separable from $\fs(\alpha,\delta,g)$.  That is, we have
assumed that the form of $R(z)$ does not depend on angular
position or $g$-band magnitude.

\section{Summary}\label{sec:sum}

In this paper we present the spectroscopic QSO catalogue from the
2dF-SDSS LRG 
and QSO (2SLAQ) Survey.  This sample is flux limited at $18<g<21.85$
(extinction corrected),and covers an area of 191.9 deg$^2$.  We have
taken new spectra of 16326 objects, of which 8764 are QSOs.  Of these, 7623
are newly discovered (the remainder were previously identified by the 2QZ and
SDSS surveys).  The full QSO sample (including objects previously
observed in the SDSS and 2QZ surveys) contains 12702 QSOs.

We present detailed completeness estimates for the survey, based on
modelling of QSO colours, including host galaxy contributions.  This
calculation shows that at $g\simeq21.85$ QSO colours are significantly
affected by the presence of a host galaxy up to $z\sim1$ in the SDSS
$ugriz$ bands.  In particular we see a significant reddening of the
objects in $g-i$ towards fainter $g$-band magnitudes.  This implies
that the host galaxies of these faint 2SLAQ QSOs are dominated by an
old ($\simgt2-3$Gyr) stellar population and not a young starburst component.

The primary aim of the 2SLAQ QSO sample is to study the luminosity and
spatial distribution of QSOs beyond the observed break in the QSO
luminosity function.  In this context it is particularly important
that detailed corrections for completeness (including host galaxy
effects) are carried out.  If we can learn how to account for the
contamination in optical samples, new larger area optical imaging
surveys (e.g. LSST; Ivezi\'{c} et al. 2008) will allow substantial
improvements in our characterization of the evolution of AGN.

The luminosity function from the final 2SLAQ catalogue is presented in
a companion paper (Croom et al. in preparation).  The 2SLAQ QSO
catalogue, along with completeness estimates, are publically available
at {\tt http://www.2slaq.info/}.

\section*{ACKNOWLEDGMENTS} 

The 2dF-SDSS LRG and QSO (2SLAQ) Survey is based on observations
carried out with the Anglo-Australian Telescope and as part of the
Sloan Digital Sky Survey. We warmly thank all the present and former
staff of the Anglo-Australian Observatory for their work in building
and operating the 2dF facility.  The 2QZ survey is based on
observations made with the Anglo-Australian Telescope and the UK
Schmidt Telescope.

Funding for the SDSS and SDSS-II has been provided by the Alfred
P. Sloan Foundation, the Participating Institutions, the National
Science Foundation, the U.S. Department of Energy, the National
Aeronautics and Space Administration, the Japanese Monbukagakusho, the
Max Planck Society, and the Higher Education Funding Council for
England. The SDSS Web Site is http://www.sdss.org/. 

The SDSS is managed by the Astrophysical Research Consortium for the
Participating Institutions. The Participating Institutions are the
American Museum of Natural History, Astrophysical Institute Potsdam,
University of Basel, University of Cambridge, Case Western Reserve
University, University of Chicago, Drexel University, Fermilab, the
Institute for Advanced Study, the Japan Participation Group, Johns
Hopkins University, the Joint Institute for Nuclear Astrophysics, the
Kavli Institute for Particle Astrophysics and Cosmology, the Korean
Scientist Group, the Chinese Academy of Sciences (LAMOST), Los Alamos
National Laboratory, the Max-Planck-Institute for Astronomy (MPIA),
the Max-Planck-Institute for Astrophysics (MPA), New Mexico State
University, Ohio State University, University of Pittsburgh,
University of Portsmouth, Princeton University, the United States
Naval Observatory, and the University of Washington. 

SMC acknowledges the support of an Australian Research Council QEII
Fellowship and a J G Russell Award from the Australian Academy of
Science.  NPR and DPS acknowledge the support of National Science
Foundation grant AST06-07634.

\appendix

\section{Observed 2SLAQ field centres}

\begin{table*}
\begin{center}
\caption{The observed 2dF field centres in the 2SLAQ survey with dates
  observed (month/year).  Also listed are a number of statistics for
  the primary sample in each field (those objects targeted on the
  first night of observation) as well as statistics including all
  targeted objects (including those configured on subsequent
  nights).  $\nobs$ is the number of 2SLAQ QSO candidates observed;
  $\nq$ is the number of good quality QSO identifications
  (identification quality 1); $f_1$ is the fraction of good quality
  identifications (quality 1) for all object types (i.e. QSOs, stars,
  galaxies etc.) as a percentage.  Note that a small number
  of fields (i.e. b00 and c00) are not in the same RA order as the
  remainder.  Also, fields d09 and d10 have the same field centres.}
\setlength{\tabcolsep}{4pt}
\begin{tabular}{cccccccccc}
\hline
Field & RA & Dec. & Date & \multicolumn{3}{c}{Primary} & \multicolumn{3}{c}{All}\\
name & (J2000) & (J2000) & (mm/yy) & $\nobs$ & $\nq$ & $f_1$ (\%) & $\nobs$ & $\nq$ & $f_1$ (\%)\\  
\hline
 a01 & 08 14 00 & -00 12 35 & 03/03  & 168 &  78 & 84.5 & 168 &  78 & 84.5\\
 a02 & 08 18 00 & -00 12 35 & 03/03  & 174 &  64 & 73.6 & 174 &  64 & 73.6\\
 a13 & 09 10 48 & -00 12 35 & 03/05  & 169 & 102 & 80.5 & 238 & 142 & 79.8\\
 a14 & 09 15 36 & -00 12 35 & 04/04  & 164 &  87 & 82.9 & 193 & 101 & 82.4\\
 a15 & 09 20 24 & -00 12 35 & 04/04  & 165 &  97 & 92.1 & 243 & 144 & 89.7\\
 a16 & 09 25 12 & -00 12 35 & 04/04  & 164 &  89 & 78.7 & 197 & 108 & 78.7\\
  "  &    "     &     "     & 03/05  & 163 &  92 & 83.4 & 215 & 123 & 84.2\\
 a17 & 09 30 00 & -00 12 35 & 03/04  & 170 &  92 & 90.0 & 238 & 133 & 89.1\\
 b00 & 10 02 00 & -00 12 35 & 04/05  & 166 &  87 & 79.5 & 226 & 123 & 79.6\\
 b01 & 10 01 00 & -00 12 35 & 04/03  & 148 &  67 & 90.5 & 148 &  67 & 90.5\\
 b02 & 10 05 00 & -00 12 35 & 03/03  & 172 &  63 & 84.3 & 172 &  63 & 84.3\\
  "  &    "     &     "     & 04/05  & 161 &  81 & 78.3 & 204 & 101 & 77.9\\
 b03 & 10 09 00 & -00 12 35 & 03/03  & 175 &  63 & 80.0 & 175 &  63 & 80.0\\
  "  &    "     &     "     & 04/05  & 161 &  84 & 82.0 & 224 & 120 & 85.3\\
 b04 & 10 13 48 & -00 12 35 & 03/04  & 175 &  80 & 86.9 & 219 & 105 & 85.8\\
 b05 & 10 18 36 & -00 12 35 & 04/04  & 165 &  74 & 79.4 & 210 &  99 & 80.0\\
 b06 & 10 23 24 & -00 12 35 & 03/04  & 168 &  54 & 69.6 & 199 &  68 & 69.8\\
  "  &    "     &     "     & 04/05  & 169 &  80 & 88.2 & 231 & 112 & 90.0\\
 b07 & 10 28 12 & -00 12 35 & 04/04  & 162 &  84 & 88.9 & 198 & 101 & 87.4\\
 b08 & 10 33 00 & -00 12 35 & 04/04  & 163 &  80 & 78.5 & 250 & 113 & 76.0\\
 b09 & 10 37 48 & -00 12 35 & 03/05  & 163 &  73 & 84.7 & 245 & 112 & 81.2\\
 b10 & 10 42 36 & -00 12 35 & 04/04  & 163 &  70 & 84.0 & 202 &  89 & 83.7\\
 b11 & 10 47 24 & -00 12 35 & 03/05  & 169 &  80 & 82.2 & 244 & 113 & 84.0\\
 b12 & 10 52 12 & -00 12 35 & 04/05  & 169 &  84 & 92.9 & 328 & 174 & 88.1\\
 b13 & 10 57 00 & -00 12 35 & 04/05  & 169 &  70 & 87.6 & 248 & 113 & 86.3\\
 b14 & 11 01 48 & -00 12 35 & 04/05  & 162 &  76 & 88.9 & 235 & 112 & 84.3\\
 b15 & 11 05 48 & -00 12 35 & 04/05  & 162 &  77 & 87.0 & 235 & 119 & 89.4\\
 b16 & 11 09 48 & -00 12 35 & 04/05  & 159 &  80 & 79.2 & 241 & 115 & 76.3\\
 c00 & 12 22 30 & -00 12 35 & 05/05  & 164 &  95 & 84.1 & 223 & 129 & 83.9\\
 c01 & 12 21 30 & -00 12 35 & 03/03  & 163 &  70 & 80.4 & 163 &  70 & 80.4\\
 c02 & 12 25 30 & -00 12 35 & 03/03  & 167 &  71 & 84.4 & 167 &  71 & 84.4\\
 c03 & 12 29 30 & -00 12 35 & 04/03  & 176 &  69 & 77.3 & 176 &  69 & 77.3\\
  "  &    "     &     "     & 04/05  & 161 &  86 & 80.1 & 161 &  86 & 80.1\\
 c04 & 12 33 30 & -00 12 35 & 04/03  & 176 &  81 & 88.1 & 176 &  81 & 88.1\\
 c05 & 12 38 18 & -00 12 35 & 03/04  & 170 &  99 & 84.1 & 198 & 119 & 85.9\\
 c06 & 12 43 06 & -00 12 35 & 04/04  & 165 &  76 & 78.8 & 223 & 107 & 78.9\\
 c07 & 12 47 54 & -00 12 35 & 03/04  & 169 &  92 & 87.0 & 215 & 116 & 86.0\\
 d03 & 13 12 00 & -00 12 35 & 04/04  & 164 &  84 & 80.5 & 164 &  84 & 80.5\\
 d04 & 13 16 48 & -00 12 35 & 03/05  & 169 &  69 & 81.7 & 252 & 117 & 82.9\\
 d05 & 13 21 36 & -00 12 35 & 04/04  & 164 &  84 & 84.1 & 207 &  91 & 73.9\\
 d06 & 13 26 24 & -00 12 35 & 03/04  & 171 & 101 & 88.9 & 217 & 124 & 87.1\\
 d07 & 13 31 12 & -00 12 35 & 04/04  & 165 &  64 & 65.5 & 201 &  82 & 67.7\\
  "  &    "     &     "     & 04/05  & 169 &  87 & 86.4 & 237 & 135 & 89.0\\
 d08 & 13 36 00 & -00 12 35 & 04/03  & 177 &  90 & 88.7 & 177 &  90 & 88.7\\
  "  &    "     &     "     & 05/05  & 159 &  94 & 89.3 & 210 & 121 & 87.6\\
 d09 & 13 40 00 & -00 12 35 & 03/03  & 176 &  69 & 79.0 & 176 &  69 & 79.0\\
 d10 & 13 40 00 & -00 12 35 & 04/03  & 177 &  97 & 88.1 & 177 &  97 & 88.1\\
  "  &    "     &     "     & 04/05  & 166 &  99 & 92.8 & 226 & 133 & 89.8\\
 d11 & 13 44 48 & -00 12 35 & 04/05  & 169 &  89 & 90.5 & 249 & 130 & 85.9\\
 d12 & 13 49 36 & -00 12 35 & 04/05  & 169 & 105 & 91.1 & 253 & 152 & 87.7\\
 d13 & 13 54 24 & -00 12 35 & 03/05  & 163 &  90 & 85.9 & 250 & 138 & 85.2\\
 d14 & 13 59 12 & -00 12 35 & 04/04  & 161 &  93 & 85.7 & 184 & 106 & 83.2\\
  "  &    "     &     "     & 04/05  & 169 &  87 & 89.3 & 232 & 131 & 90.5\\
 d15 & 14 04 00 & -00 12 35 & 03/05  & 163 &  83 & 81.0 & 229 & 127 & 82.1\\
 d16 & 14 08 48 & -00 12 35 & 04/04  & 163 &  93 & 88.3 & 250 & 141 & 86.4\\
 d17 & 14 13 36 & -00 12 35 & 03/05  & 169 &  73 & 79.9 & 229 & 104 & 80.3\\
 \hline
\label{tab:fields1}
\end{tabular}
\end{center}
\end{table*}

\addtocounter{table}{-1}

\begin{table*}
\begin{center}
\caption{{\it continued.}}
\setlength{\tabcolsep}{4pt}
\begin{tabular}{cccccccccc}
\hline
Field & RA & Dec. & Date & \multicolumn{3}{c}{Primary} & \multicolumn{3}{c}{All}\\
name & (J2000) & (J2000) & (mm/yy) & $\nobs$ & $\nq$ & $f_1$ (\%) & $\nobs$ & $\nq$ & $f_1$ (\%)\\  
\hline
 e01 & 14 34 00 & -00 12 35 & 04/03  & 176 &  73 & 80.7 & 176 &  73 & 80.7\\
  "  &    "     &     "     & 04/05  & 169 &  94 & 72.2 & 169 &  94 & 72.2\\
 e02 & 14 38 00 & -00 12 35 & 04/03  & 176 &  73 & 79.5 & 176 &  73 & 79.5\\
  "  &    "     &     "     & 04/05  & 160 &  87 & 85.0 & 160 &  87 & 85.0\\
 e03 & 14 42 48 & -00 12 35 & 04/05  & 162 &  90 & 88.3 & 253 & 148 & 86.6\\
 e04 & 14 47 36 & -00 12 35 & 04/04  & 164 &  82 & 73.2 & 194 &  97 & 71.6\\
  "  &    "     &     "     & 05/05  & 166 &  95 & 86.1 & 231 & 132 & 85.3\\
 e05 & 14 52 24 & -00 12 35 & 04/05  & 162 &  84 & 79.6 & 222 & 123 & 81.5\\
 e06 & 14 57 12 & -00 12 35 & 04/04  & 161 &  76 & 75.2 & 161 &  76 & 75.2\\
  "  &    "     &     "     & 04/05  & 161 &  84 & 81.4 & 229 & 131 & 83.8\\
 e07 & 15 02 00 & -00 12 35 & 07/05  & 159 &  85 & 78.0 & 257 & 133 & 75.1\\
 e08 & 15 06 48 & -00 12 35 & 04/05  & 161 &  79 & 86.3 & 161 &  79 & 86.3\\
  "  &    "     &     "     & 08/05  & 159 &  59 & 62.3 & 217 & 102 & 70.5\\
 e09 & 15 11 36 & -00 12 35 & 07/05  & 164 &  71 & 76.2 & 240 &  96 & 67.5\\
 e10 & 15 16 24 & -00 12 35 & 08/05  & 164 &  65 & 64.0 & 198 &  87 & 68.7\\
 s06 & 21 21 36 & -00 15 00 & 09/03  & 172 &  60 & 80.8 & 172 &  60 & 80.8\\
 s07 & 21 26 24 & -00 15 00 & 10/04  & 168 &  56 & 67.9 & 243 &  88 & 66.7\\
  "  &    "     &     "     & 07/05  & 159 &  61 & 69.2 & 207 &  93 & 73.9\\
 s08 & 21 31 12 & -00 15 00 & 10/04  & 171 &  54 & 80.1 & 222 &  67 & 74.3\\
 s09 & 21 36 00 & -00 15 00 & 10/04  & 168 &  32 & 39.3 & 168 &  32 & 39.3\\
  "  &    "     &     "     & 08/05  & 159 &  47 & 72.3 & 196 &  63 & 73.5\\
 s10 & 21 40 48 & -00 15 00 & 07/05  & 165 &  60 & 72.1 & 224 &  84 & 71.4\\
 s11 & 21 45 36 & -00 15 00 & 07/05  & 164 &  61 & 72.0 & 224 &  82 & 68.3\\
 s12 & 21 50 24 & -00 15 00 & 08/03  & 176 &  45 & 71.0 & 176 &  45 & 71.0\\
 s25 & 22 52 48 & -00 15 00 & 08/03  & 176 &  87 & 85.2 & 176 &  87 & 85.2\\
 s26 & 22 57 36 & -00 15 00 & 10/04  & 150 &  59 & 61.3 & 150 &  59 & 61.3\\
  "  &    "     &     "     & 08/05  & 164 &  81 & 68.3 & 164 &  81 & 68.3\\
 s27 & 23 02 24 & -00 15 00 & 09/03  & 175 &  82 & 80.0 & 175 &  82 & 80.0\\
 s28 & 23 07 12 & -00 15 00 & 10/04  & 150 &  74 & 74.7 & 176 &  93 & 75.6\\
 s29 & 23 12 00 & -00 15 00 & 10/04  & 156 &  81 & 88.5 & 215 & 112 & 89.8\\
 s30 & 23 16 48 & -00 15 00 & 10/04  & 150 &  84 & 90.0 & 212 & 113 & 88.2\\
 s31 & 23 21 36 & -00 15 00 & 07/05  & 159 &  89 & 84.3 & 208 & 113 & 83.7\\
 s32 & 23 26 24 & -00 15 00 & 08/05  & 159 &  94 & 86.2 & 198 & 117 & 87.4\\
 s47 & 00 38 24 & -00 15 00 & 08/05  & 164 &  86 & 80.5 & 234 & 131 & 79.5\\
 s48 & 00 43 12 & -00 15 00 & 10/04  & 171 &  86 & 83.0 & 271 & 145 & 82.7\\
 s49 & 00 48 00 & -00 15 00 & 10/04  & 168 &  69 & 60.7 & 168 &  69 & 60.7\\
 s50 & 00 52 48 & -00 15 00 & 08/03  & 174 &  76 & 84.5 & 174 &  76 & 84.5\\
 s51 & 00 57 36 & -00 15 00 & 10/04  & 167 &  79 & 85.0 & 228 & 108 & 85.1\\
 s52 & 01 02 24 & -00 15 00 & 09/03  & 172 &  72 & 79.7 & 172 &  72 & 79.7\\
 s67 & 02 14 24 & -00 15 00 & 10/04  & 156 &  72 & 82.7 & 241 & 100 & 74.3\\
 s68 & 02 19 12 & -00 15 00 & 10/04  & 150 &  51 & 53.3 & 150 &  51 & 53.3\\
 s69 & 02 24 00 & -00 15 00 & 09/03  & 173 &  52 & 64.7 & 173 &  52 & 64.7\\
 s70 & 02 28 48 & -00 15 00 & 10/04  & 150 &  63 & 83.3 & 213 &  90 & 77.5\\
\hline
\end{tabular}
\end{center}
\end{table*}


\vspace{1.0truecm}
This paper has been produced using the Blackwell Scientific Publications 
\TeX macros.

\end{document}